\documentclass[sigconf]{acmart} 

\AtBeginDocument{%
  \providecommand\BibTeX{{%
    \normalfont B\kern-0.5em{\scshape i\kern-0.25em b}\kern-0.8em\TeX}}}

\copyrightyear{2022}
\acmYear{2022}
\setcopyright{acmcopyright}\acmConference[CHI '22]{CHI Conference on Human
Factors in Computing Systems}{April 29-May 5, 2022}{New Orleans, LA, USA}
\acmBooktitle{CHI Conference on Human Factors in Computing Systems (CHI '22),
April 29-May 5, 2022, New Orleans, LA, USA}
\acmPrice{15.00}
\acmDOI{10.1145/3491102.3517455}
\acmISBN{978-1-4503-9157-3/22/04} 

\usepackage{soul}  
\usepackage{color} 
\usepackage{xspace}
\usepackage{listings}

\newcommand{\ie}{{i.e.,}\xspace}
\newcommand{\eg}{{e.g.,}\xspace}
\newcommand{\cf}{{c.f.}\xspace}
\newcommand{\ea}{{et~al.}\xspace}

\newcommand{\etc}{{etc.}\xspace}
\newcommand{\bstart}[1]{\vspace{1.5mm} \noindent{\textbf{#1:}}}
\newcommand{\bstartnc}[1]{\vspace{1.5mm} \noindent{\textbf{#1}}}

\definecolor{lightorange}{rgb}{1,0.8,0.4}
\definecolor{lightorange}{RGB}{230, 170, 50}
\definecolor{lightgreen}{RGB}{121,210,121}
\definecolor{lightteal}{RGB}{121,199,210}
\definecolor{lightblue}{RGB}{100,212,239}
\definecolor{lightpurple}{RGB}{153,102,255}
\definecolor{lightred}{RGB}{245, 132, 120}
\definecolor{red}{RGB}{178,34,34}
\definecolor{revred}{RGB}{245,25,25}
\definecolor{gray}{RGB}{166,166,166}
\definecolor{indexBlue}{cmyk}{0.9,0.8,0,0}
\definecolor{indexGreen}{cmyk}{0.8,0.2,0.8,0.55}
\definecolor{deepblue}{cmyk}{0.9,0.75,0,0.5}
\definecolor{deepred}{cmyk}{0,0.75,0.75,0.4}
\definecolor{pink}{RGB}{245,61,144}
\definecolor{pink}{RGB}{214, 114, 0}

\definecolor{codeOrange}{cmyk}{0,0.73,1,0}
\definecolor{codeGreen}{cmyk}{0.85,0.2,1,0.15}

\lstdefinelanguage{JavaScript}{
  keywords={typeof, new, catch, function, return, null, catch, switch, var, if, in, while, do, else, case, break},
  keywordstyle=\color{blue},
  ndkeywords={class, export, boolean, throw, implements, import, this},
  ndkeywordstyle=\color{darkgray}\bfseries,
  identifierstyle=\color{codeGreen},
  classoffset = 1,
  morekeywords={true, false},
  keywordstyle=\color{codeOrange},
  classoffset = 2,
  morekeywords={0,1,2,3,4,5,6,7,8,9},
  keywordstyle=\color{codeOrange},
  sensitive=false,
  comment=[l]{//},
  morecomment=[s]{/*}{*/},
  commentstyle=\color{purple}\ttfamily,
  stringstyle=\color{codeOrange}\ttfamily,
  numberstyle=\color{codeOrange}\ttfamily,
  morestring=[b]',
  morestring=[b]",
  literate=%
    {0}{{{\color{codeOrange}0}}}1
    {1}{{{\color{codeOrange}1}}}1
    {2}{{{\color{codeOrange}2}}}1
    {3}{{{\color{codeOrange}3}}}1
    {4}{{{\color{codeOrange}4}}}1
    {5}{{{\color{codeOrange}5}}}1
    {6}{{{\color{codeOrange}6}}}1
    {7}{{{\color{codeOrange}7}}}1
    {8}{{{\color{codeOrange}8}}}1
    {9}{{{\color{codeOrange}9}}}1
}

\newcommand{\cicero}[1]{\textcolor{deepblue}{\texttt{#1}}}
\newcommand{\datafield}[1]{\textcolor{deepred}{\texttt{#1}}}
\newcommand{\specin}[1]{\lstinline!#1!}
\newcommand{\asp}[1]{\textcolor{deepblue}{\texttt{#1}}}
\newcommand{\claim}[1]{{\textbf{#1}}}

\newcounter{dcIndex}
\newcommand{\consideration}[2]{\vspace{2mm} \noindent\refstepcounter{dcIndex}\label{#2}\textbf{\textcolor{indexBlue}{(D\thedcIndex)} #1}}
\newcounter{pcIndex}
\newcommand{\principle}[2]{\refstepcounter{pcIndex}\label{#2}\textbf{#1 \textcolor{indexGreen}{(P\thepcIndex)}}}

\newcommand{\cRef}[1]{\textbf{\textcolor{indexBlue}{D\ref{#1}}}}
\newcommand{\pRef}[1]{\textbf{\textcolor{indexGreen}{P\ref{#1}}}}

\AtBeginMaketitle{%
\let\oldAtBeginDocument\AtBeginDocument%
\renewcommand\AtBeginDocument[1]{#1}
\setcopyright{acmcopyright}
\copyrightyear{2022}
\acmYear{2022}
\acmDOI{10.1145/3491102.3517455}
\acmISBN{978-1-4503-9157-3/22/04}
\let\AtBeginDocument\oldAtBeginDocument%
}

\begin{document}


\title{Cicero: A Declarative Grammar for Responsive Visualization}

\settopmatter{authorsperrow=4}
\author{Hyeok Kim}
\affiliation{%
  \institution{Northwestern University}
  \city{Evanston}
  \state{IL}
  \country{U.S.A.}}
\email{hyeok@northwestern.edu}

\author{Ryan A. Rossi}
\affiliation{%
  \institution{Adobe Research}
  \city{San Jose}
  \state{CA}
  \country{U.S.A.}}
\email{rrossi@adobe.com}

\author{Fan Du}
\affiliation{%
  \institution{Adobe Research}
  \city{San Jose}
  \state{CA}
  \country{U.S.A.}}
\email{fdu@adobe.com}

\author{Eunyee Koh}
\affiliation{%
  \institution{Adobe Research}
  \city{San Jose}
  \state{CA}
  \country{U.S.A.}}
\email{eunyee@adobe.com}

\author{Shunan Guo}
\affiliation{%
  \institution{Adobe Research}
  \city{San Jose}
  \state{CA}
  \country{U.S.A.}}
\email{sguo@adobe.com}

\author{Jessica Hullman}
\affiliation{%
  \institution{Northwestern University}
  \city{Evanston}
  \state{IL}
  \country{U.S.A.}}
\email{jhullman@northwestern.edu}

\author{Jane Hoffswell}
\affiliation{%
  \institution{Adobe Research}
  \city{Seattle}
  \state{WA}
  \country{U.S.A.}}
\email{jhoffs@adobe.com}

\renewcommand{\shortauthors}{Kim, et al.}



\begin{abstract}
\noindent 
Designing responsive visualizations can be cast as applying transformations to a source view to render it suitable for a different screen size.
However, designing responsive visualizations is often tedious as authors must manually apply and reason about candidate transformations.
We present Cicero, a declarative grammar for concisely specifying responsive visualization transformations which paves the way for more intelligent responsive visualization authoring tools.
Cicero's flexible \emph{specifier} syntax allows authors to select visualization elements to transform, independent of the source view's structure.
Cicero encodes a concise set of \emph{actions} to encode a diverse set of transformations in both desktop-first and mobile-first design processes. 
Authors can ultimately reuse design-agnostic transformations across different visualizations.
To demonstrate the utility of Cicero, we develop a compiler to an extended version of Vega-Lite, and provide principles for our compiler. 
We further discuss the incorporation of Cicero into responsive visualization authoring tools, such as a design recommender. 
\end{abstract}










\begin{teaserfigure}
  \includegraphics[width=\textwidth]{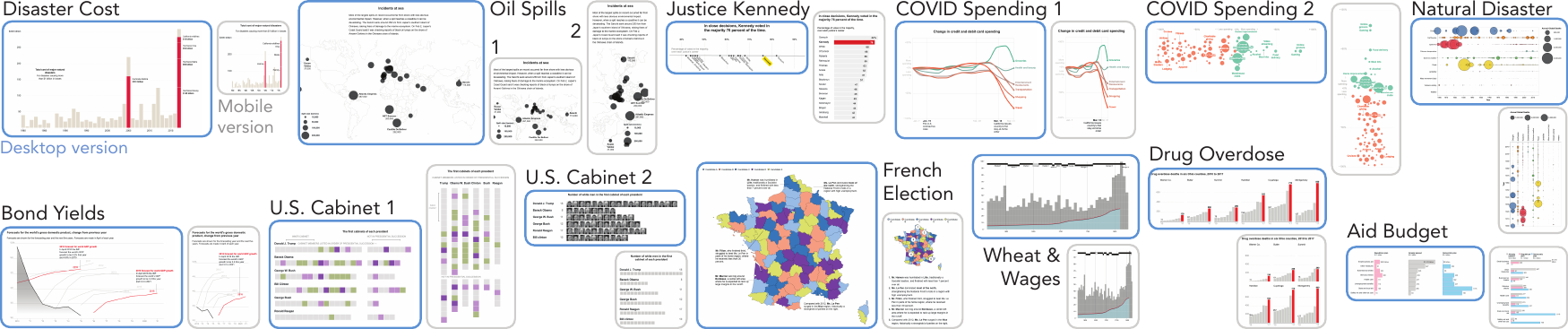}
  \caption{Thirteen responsive visualization use cases reproduced using Cicero. 
  The blue- and gray-bordered views are the desktop and mobile versions, respectively. The mobile versions of the Oil Spills case are from (1)~the original article and (2)~the version suggested by Hoffswell~\ea~\cite{hoffswell:responsive2020}. Full size images are included in the Supplemental Material (\url{https://osf.io/eg4xq}).}
  \Description{This figure includes 13 responsive visualization use cases reproduced using Cicero. Each case consists of a pair of the desktop and mobile versions. The cases are: Disaster Cost (a bar chart), Oil Spills (a map with sized circles), Justice Kennedy (a dot plot in the desktop version and bar chart in the mobile version), COVID Spending 1 (a line chart), COVID Spending 2 (a bubble chart), Natural Disaster (a bubble chart), Bond Yields (a line chart), U.S. Cabinet 1 (an icon array), U.S. Cabinet 2 (an icon array), French Election (choropleth), Wheat and Wages (a bar and area chart), Drug Overdose (small multiples of bar charts), and Aid Budget (a grouped bar chart).}
  \label{fig:teaser}
\end{teaserfigure}

\maketitle


\section{Introduction}
\begin{figure}[t]
    \centering
    \includegraphics[width=\columnwidth]{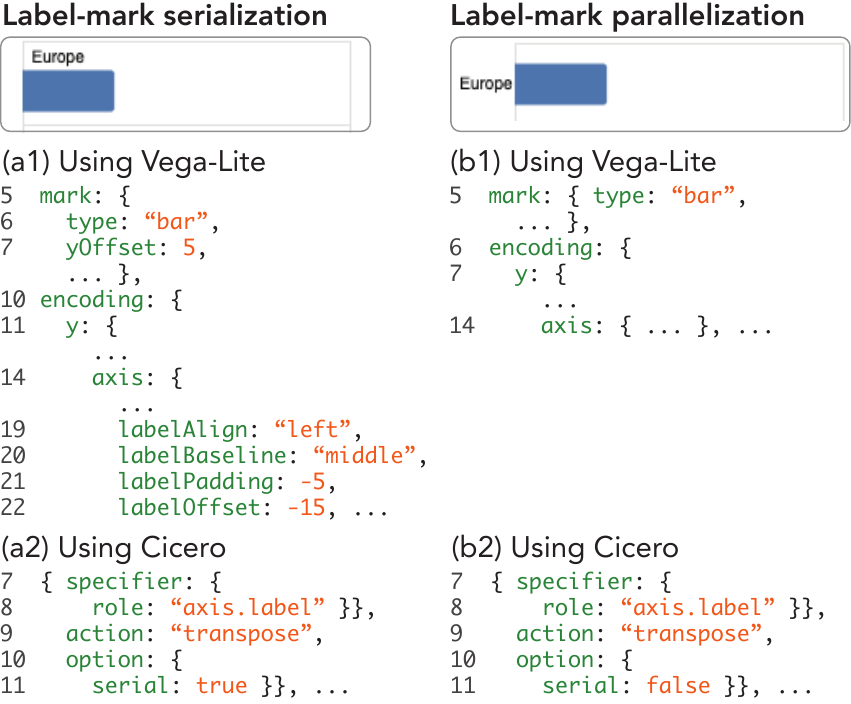}
    \caption{Design specifications for label-mark serialization using (a1) Vega-Lite and (a2) Cicero and parallelization using (b1) Vega-Lite and (b2) Cicero.}
    \Description{The left column is for label-mark serialization and the right column is for label-mark parallelization. Each column consists of the demonstration image of a bar chart, a Vega-Lite code snippet, and a Cicero code snippet. On the left column, the bar chart has a label vertically higher to a bar mark; the code snippet, a1, is serialization using Vega-Lite; and the code snippet, a2, is serialization using Cicero. On the right column, the bar chart has a label and horizontally left to a bar mark; the code snippet, b1, is parallelization using Vega-Lite; and the code snippet, b2, is parallelization using Cicero.}
    \label{fig:vlComparison}
\end{figure}

Responsive visualizations adapt visualization content for different screen types, making them essential for most Web-based contexts due to an increasing proportion of mobile viewers.
Responsive visualization authoring environments, however, tend to require considerable manual effort on the part of visualization designers. 
Prior findings on responsive visualization design practices~\cite{hoffswell:responsive2020,kim:responsive2021} indicate that authors often start from a source view and then apply responsive transformations to produce a set of target views optimized for different screen types.
However, this approach can be tedious as authors must manually explore, apply, and evaluate different responsive strategies one by one.
For example, authors might create responsive views by crafting an artboard and/or specification per responsive view, which is particularly problematic when one of the responsive views is revised.
They may have difficulty in expressing changes that occur across a design specification (\eg~example cases in \autoref{fig:vlComparison} and \autoref{fig:flexibility}).
Authoring painpoints like these suggest a need for more intelligent authoring tools, such as semi- or fully automated  recommenders that support exploring and reasoning about responsive design strategies~\cite{kim:responsive2021,kim:insight2021}.

A key step toward such intelligent responsive visualization authoring tools is a concise, declarative grammar that can express a diverse set of transformation strategies.
While declarative visualization grammars like Vega~\cite{satyanarayan:vega2016} and Vega-Lite~\cite{satyanarayan:vega-lite2017} are well suited to developing more sophisticated visualization authoring tools, they are not necessarily well suited to representing visualization transformations; 
Hoffswell~\ea~\cite{hoffswell:responsive2020} observe that different edit properties for text and marks in Vega-Lite~\cite{satyanarayan:vega-lite2017} make it complicated to create the specifications for multiple versions of a visualization despite its high expressiveness. 
Indeed, many responsive visualization strategies that researchers have identified~\cite{kim:responsive2021} can be written in Vega-Lite with high complexity. 
For instance, serializing labels and marks using Vega-Lite (\ie~placing them in a vertical order~\cite{kim:responsive2021}) requires layout adjustment keywords (\autoref{fig:vlComparison}a1, line 7, 19--22), while parallelizing them (\ie~arraying them horizontally) does not require layout modifications in Vega-Lite (\autoref{fig:vlComparison}b1, line 14).
Whereas Vega-Lite requires authors to create separate specifications for each responsive view that interleave complex layout changes throughout the specifications, a declarative grammar for responsive transformations can express the same strategies in a simpler way as shown in \autoref{fig:vlComparison} (a2) and (b2).
Such an approach can help visualization authors easily and quickly compose responsive design specifications and can help developers to more effectively develop authoring tools for responsive visualization.


To this end, we present Cicero: a flexible, expressive, and reusable declarative grammar for specifying responsive visualization transformations.
The flexible \textit{specifier} syntax of Cicero enables querying visualization elements using their role (\eg~mark, axis labels, title), 
underlying data, and attributes of visualization elements, independent of the structure of a source view specification. 
Cicero provides a compact set of \textit{action} predicates (\cicero{add}, \cicero{duplicate}, \cicero{remove}, \cicero{replace}, \cicero{swap}, \cicero{modify}, \cicero{reposition}, and \cicero{transpose})
that can encode a diverse range of transformation techniques (\autoref{fig:cicero_overview}c). 
Moreover, Cicero supports extracting and reusing generalizable transformations strategies across multiple responsive specifications. 
For example, the expressions (a2) and (b2) in \autoref{fig:vlComparison} can be reused on other visualizations with bar-like marks and axis labels. 

To demonstrate the utility of Cicero, we develop a Cicero compiler for an extended version of Vega-Lite that we adapted to support annotations and other narrative devices and reproduce 13 real-world examples in Cicero (\autoref{sec:examples}).
We provide a set of principles for developing our Cicero compiler in terms of desirable properties of the association of visualization elements, preferable default behavior, and how to manage conflicts between transformations (\autoref{sec:compiler}). 
As Cicero is agnostic to the underlying structure of a source visualization, it can be leveraged in different visualization authoring tools.
To demonstrate the feasibility of Cicero in such authoring tools, we describe how Cicero applies to a prototype recommender we developed for responsive transformations as a proof of concept and envision an approach to mixed-initiative authoring tools (\autoref{sec:recommender}).
Future work can implement a Cicero compiler for other declarative grammars like the original Vega-Lite~\cite{satyanarayan:vega-lite2017} or ggplot2~\cite{wickham:ggplot22010} and other recommender approaches (\eg~\cite{wu:mobilevisfixer2020}). 





\section{Related Work}
This work is motivated by prior research on responsive visualization and declarative visualization grammars.

\subsection{Responsive Visualization}
Prior research has examined how visualization authors customize a visualization for smaller screens in terms of visual elements and structure~\cite{wu:visizer2012,giacomo:retargeting2015,andrews:responsive2018}, and interaction methods~\cite{jakobsen:2013}.
For instance, VISizer~\cite{wu:visizer2012} provides a point-of-interest-based framework to resize a visualization while preserving regions with important insights. 
Recent works~\cite{hoffswell:responsive2020,kim:responsive2021} provide a more comprehensive snapshot of current responsive visualization design practices.
Motivated by a qualitative analysis of 231 responsive visualizations and a formative interview study, Hoffswell~\ea~\cite{hoffswell:responsive2020} implement an authoring tool that supports editing across different responsive views via simultaneous previews and global edits, as well as view-specific customization. 
Using a similar approach, Kim~\ea~\cite{kim:responsive2021} present a set of responsive visualization design patterns and identify a trade-off between achieving appropriate graphical density for each view and preserving intended takeaways across transformations. 
To address the trade-off between density and takeaways, Kim~\ea~\cite{kim:insight2021} provide a set of task-oriented insight preservation measures for a responsive visualization recommender limited to a small set of design transformations (\eg~aggregation, axes-transposing). 
A recent machine learning-based approach~\cite{wu:autolayout2021,wu:mobilevisfixer2020} provides automated methods to configure visualization layouts based on the chart size using a set of simple heuristics, yet it does not offer a grammar that can express a large set of responsive visualization techniques. 

Responsive design has been well-studied for the Web more generally~\cite{bryant:responsiveWeb2012,mohorovicic:responsiveWeb2013}, but such techniques are not directly applicable to responsive visualization design because they are intended for Web layouts and based on limited knowledge of visualization design.
For example, CSS media queries~\cite{cssmediaquery} express breakpoints for each responsive version of the contents.
CSS specifications under a media query of \texttt{@media screen and max-width 600px} are shown only on a screen-side application (\eg~Web browser) with width $\leq$ 600px.
Similarly, CSS specifications under a media query of \texttt{@media speech} are used by speech synthesizers like a screen reader.
However, using CSS alone cannot enable specification of many responsive transformations specific to visualization, such as transposing axis (requiring changes to scale functions), un-fixing tooltip positions, changing mark types (requiring dynamic positioning), and transforming data (requiring custom JavaScript functions). 

In practice, designers create responsive visualizations with multiple tools in an iterative manner. 
D3.js~\cite{bostock:d32011} is a highly expressive JavaScript (JS) library for SVG- or Canvas-based visualizations. 
According to prior work on visualization authoring practices~\cite{bently:2021,sam:ai2htmlIndustry2018}, designers often use D3.js (or equivalent tools) with ai2html~\cite{ai2html}, which renders Adobe Illustrator vector images (\texttt{.ai} files) to HTML. 
Designers first draw a visualization using D3.js~\cite{bostock:d32011}, then load and edit the SVG graphic of the visualization as responsive `artboards' in Adobe Illustrator~\cite{sam:ai2htmlIndustry2018}. 
Authors can also define responsive condition parameters for interactive visualizations using D3.js (\eg~scale functions for \textit{x} and \textit{y} positions to be swapped for mobile screen). 
R3S.js~\cite{leclaire:r3sjs2015} offers programming interfaces for such parameterization by extending D3.js~\cite{bostock:d32011}. 
However, it is not fully declarative, so authors need to imperatively define each transformation, which requires programming expertise.
For example, to reposition a tooltip, which is a common responsive transformation strategy~\cite{kim:responsive2021}, R3S.js requires the use of custom CSS rules and/or JS functions.

For simple charts and quick edits, authors can utilize responsive properties of existing tools like Vega, Google Chart, and Microsoft Power BI. 
While Vega~\cite{satyanarayan:vega2016} and Vega-Lite~\cite{satyanarayan:vega-lite2017} support some `sensible' defaults, such as fitting the number of axis labels to the chart size, users need to have fully defined specifications for each of the responsive views. 
Google Chart~\cite{googlechart} offers several default settings for mobile views such as truncating labels with an ellipsis~(...). 
Power BI~\cite{powerbi} provides defaults for responsiveness (\eg~making a visualization scrollable, rearranging legends, removing axis, \etc)~\cite{gal:powerbi2017}. 
While these tools can simplify the design process, their limited expressiveness may prevent authors from specifying intended responsive transformations, limiting their ability to convey insights. 

Lastly, commercial tools like ZingChart and DataWrapper allow for responsive settings. 
ZingChart~\cite{zingchart} provides `media rules’ through which a designer can declare a screen size condition for a visualization element (\eg~label: `October 4' for screen \mbox{size $>$ 500} and `Oct. 4' for screen \mbox{size $<$ 500}). 
However, those media rules are dependent on the chart type---for example, transposing a scatterplot and a bar chart requires changes to data structure and the chart type, respectively---which limits the expressiveness and flexibility for responsive transformations.
DataWrapper~\cite{datawrapper}, an authoring tool for communicative visualizations, allows authors to choose whether and how to show a visualization element for mobile screens (\eg~showing a table as a stack of cards~\cite{rost:datawrapperTable2020}, or numbering annotations~\cite{rost:datawrapperAnno2020}). 
However, it is not available in the form of a declarative grammar which limits how easily it can be extended or applied to future authoring tools, such as a mixed-initiative authoring tool.


\subsection{Declarative Visualization Grammars}

Declarative grammars help visualization authors to avoid complex programming through a compiler that implements user-declared specifications (\eg~\cite{satyanarayan:vega2016,satyanarayan:vega-lite2017,moritz:draco2019,wickham:ggplot22010,kim:gemini2021,hoffswell:setcola2018}). 
For example, a Vega-Lite~\cite{satyanarayan:vega-lite2017} specification uses JavaScript object notation (JSON) to encode chart size, data source and transformation, visual encodings, multiple views, and user interactions using predefined primitives. 
Some declarative grammars target specific use-cases by leveraging more general-purpose grammars. 
For example, Gemini’s animated transition grammar formalizes chart animation entities~\cite{kim:gemini2021} based on starting and ending visualizations specified using Vega~\cite{satyanarayan:vega2016}. 
Moreover, declarative grammars facilitate computational operations on visualization specification, which enables the development of useful visualization applications on top of the underlying grammar. 
For example, many end-user tools like visualization recommender(s)~\cite{wongsuphasawat:voyager2016,wongsuphasawat:voyager22017,moritz:draco2019} and editor(s)~\cite{Satyanarayan:lyra2014} use Vega-Lite~\cite{satyanarayan:vega-lite2017} to represent the visualization design specification.
In responsive visualization settings, Hoffswell~\ea~\cite{hoffswell:responsive2020} provide a design editor using Vega-Lite~\cite{satyanarayan:vega-lite2017}, and Kim~\ea~\cite{kim:insight2021} propose automated recommendation of responsive visualization designs using Draco~\cite{moritz:draco2019}.


However, existing declarative visualization grammars are often limited when it comes to supporting expressive responsive visualization designs. 
For example, common responsive visualization strategies like fixing a tooltip position, aggregation, internalizing labels, and externalizing annotations (\cf~\cite{kim:responsive2021}) are not supported or are complicated to specify in Vega-Lite~\cite{satyanarayan:vega-lite2017}. 
In addition, many commonly used visualization grammars (\eg~ggplot2~\cite{wickham:ggplot22010}, Vega-Lite~\cite{satyanarayan:vega-lite2017}) require authors to define multiple full visualization specifications for each responsive view, which makes it difficult to propagate changes from one design to another.
ZingChart~\cite{zingchart} provides `media rules’ to specify conditions for responsive properties, yet it is often difficult (or impossible) to express a large set of design transformations like transposing layout or changing mark types.

Our approach proposes a novel declarative grammar that can express various responsive transformations, accompanied by a compiler built on an extended version of Vega-Lite.
To demonstrate the utility of Cicero for visualization tooling, we develop a proof-of-concept prototype recommender for responsive design transformations that encodes a larger set of design strategies than the scope of Kim~\ea~\cite{kim:insight2021}, using Cicero as the representation method.




\section{Three Design Guidelines for a Responsive Visualization Grammar}
\noindent We derive three central design guidelines for a responsive visualization grammar based on prior work~\cite{hoffswell:responsive2020,kim:responsive2021,sam:ai2htmlIndustry2018,zingchart,bently:2021,leclaire:r3sjs2015,andrews:responsive2018}.

\consideration{Be expressive.}{dc:expressive}
A responsive visualization grammar should be able to express a diverse set of responsive design strategies spanning different visualization elements.
One approach is to characterize a responsive transformation strategy as a tuple of the visualization element(s) to change and a transformation action~\cite{hoffswell:responsive2020,kim:responsive2021}.
Selecting visualization element(s) should support varying levels of customization for responsive transformations because transformations can include both systematic changes (\eg~externalizing all text annotations or shortening axis labels) and individual changes (\eg~externalizing a subset of annotations or highlighting a particular mark)~\cite{kim:responsive2021}.
A grammar needs to express responsive transformations as a concise set of `actions' describing how visualization elements are changed~\mbox{\cite{kim:responsive2021,hoffswell:responsive2020}.}
\claim{To be expressive, our grammar provides (1) a query syntax for selecting visualization elements both systematically and individually and (2) consistent, high-level action predicates that can encode a diverse set of responsive design strategies.}

\begin{figure}[t]
    \centering
    \includegraphics[width=\columnwidth]{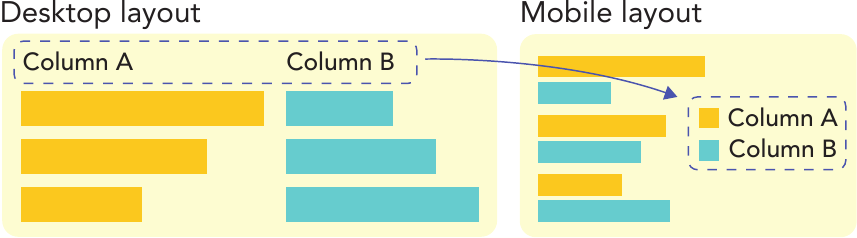}
    \caption{Responsive transformation from axis labels to a legend accompanied by a layout change for smaller display.}
    \Description{This figure has two sections. The left section shows the desktop layout of a grouped bar chart with two column groups for Column A and Column B. This bar chart has column labels, and the bars for Column A are colored as orange, and those for Column B are colored as blue. The right section shows the mobile layout where those two groups are converted to row groups. The column labels are also converted to a nominal color legend.}
    \label{fig:flexibility}
\end{figure}

\consideration{Be flexible.}{dc:flexible}
A responsive visualization grammar should offer flexibility in how an author can specify the behavior of an entity under a responsive transformation, independent of how the entity is expressed in the specification (or structure) of the source visualization. 
For example, suppose a visualization that has a nominal color encoding that maps dog, cat, and fox to red, blue, and green.
Then, to select red marks, some authors can specify simply ``red marks'' (using attribute) while others can make the same selection by specifying ``marks for dog'' (using data).
Furthermore, responsive transformations can occur across different visualization elements.
For instance, as illustrated in \autoref{fig:flexibility}, one can change the layout by moving a column element to the row (partial view transpose) to accommodate a portrait aspect ratio.
Following the previous transformation, the column labels can be replaced with a legend if there is a redundant mark property encoding.
\claim{To be flexible, our responsive visualization grammar supports multiple expressions for specifying visualization elements that can be independent of the structure of a visualization.} 


\consideration{Be reusable.}{dc:reusable}
A responsive visualization grammar should enable authors to easily (\ie~without making big changes) reapply generalizable responsive transformations across different visualizations.
While reuse is straightforward for visualizations sharing the same properties, many responsive designs utilize generic transformations that are independent of the specific chart design, data, or base visualization (\eg~transposing the layout, numbering annotations, using a fixed tooltip position).
Moreover, authors might want to repeat techniques only for certain features of a visualization (\eg~removing a data field regardless of chart type).
\claim{To be reusable, our responsive visualization grammar represents each responsive transformation in a form that helps users to easily extract and apply transformations to other views.} 

\begin{figure}[t]
    \centering
    \includegraphics[width=\columnwidth]{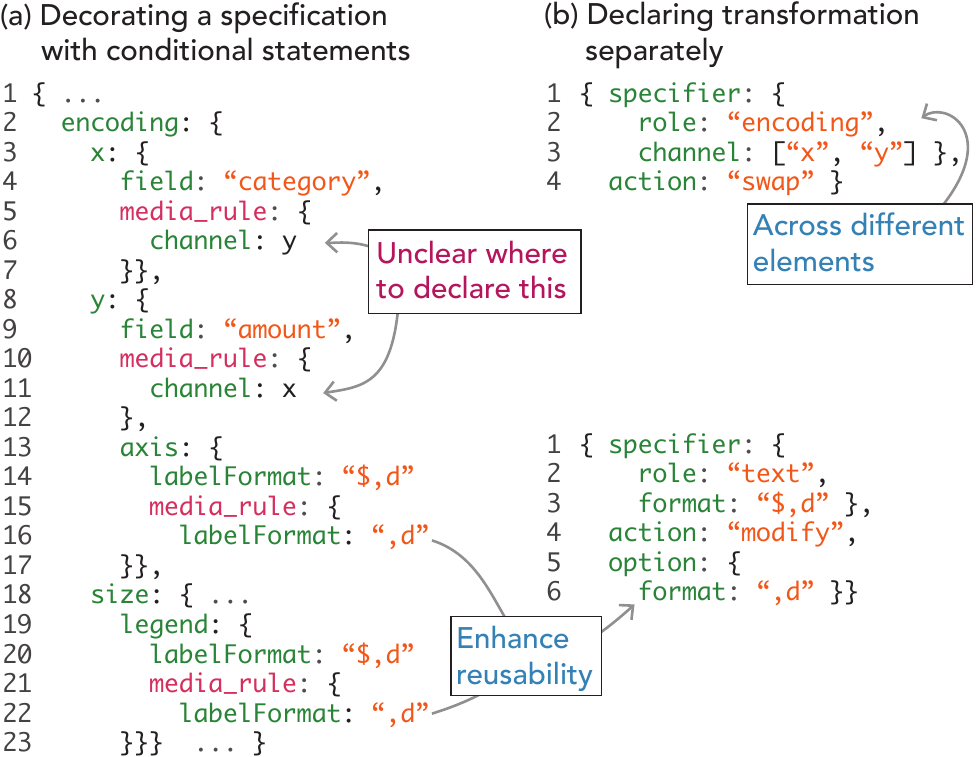}
    \caption{Two possible approaches to specifying responsive transformations. (a) Decorating a specification with conditional statements. (b) Separately defining responsive transformations.}
    \Description{This figure has two sections. The left section, A, has a hypothetical code snippet that encodes responsive transformations as decorating statements to a visualization design specification. Line five to seven and line ten to twelve in the left section indicate the lack of clarity about where to declare such a decorating statement because the transformation is about two different elements. Line fifteen to seventeen and line twenty-one to twenty-three in the left section repeat the same decorating statement. The right section, B, has a hypothetical code snippet using a separate specification for two transformations. The first rule in the right section uses a single rule that allows specifying changes across different elements. The second rule in the right section uses a single rule for the repeated transformations in the left section, featuring enhanced reusability.}
    \label{fig:approaches}
\end{figure}

With these guidelines in mind, there are several possible approaches for specifying responsive transformations, such as: (1)~decorating a complete visualization specification and (2)~separately defining responsive transformations. 
The first approach uses conditional keywords (\eg~\texttt{media\_rule} in ZingChart~\cite{zingchart}) to express transformations.
For example, in \autoref{fig:approaches}a, the \texttt{media\_rule} keywords for the \textit{x} (line 5--7) and \textit{y} (line 10--12) encodings describe the changes for each encoding when viewed in a media format (\eg~a `swap' action).
The \texttt{media\_rule} keywords for the \textit{y} \texttt{axis} (line 15--17) and the  \textit{size} \texttt{legend} (line 21--22) describe the same change to the label format for both types of elements.
For the same set of transformations, the second approach in \autoref{fig:approaches}b directly declares that the two position channels should be swapped and concisely describes changes to the label format for all text elements.
While we choose to use the JSON format, other formats could be used to extend our approach; for example, Altair~\cite{VanderPlas:altair2018} is a Python wrapper for Vega-Lite~\cite{satyanarayan:vega-lite2017} that leverages object-method chains rather than Vega-Lite's JSON format.

While the first approach simplifies the learning process by extending an existing grammar, it can sometimes be tedious and unclear how to specify responsive transformations that apply to multiple elements.
In particular, this approach often requires a single responsive change (e.g., transposing an axis) to be interleaved across multiple parts of the specification (\autoref{fig:approaches}a, Line 5--7 and 10--12).
In contrast, the second approach can enhance the reusability (\cRef{dc:reusable}: reusable) of a transformation specification by separating the desired responsive changes from the original visualization design.
Furthermore, this approach can support more generalizable transformations that are independent of the original visualization structure (\cRef{dc:flexible}: flexible; \eg~changing all text formats directly).
Therefore, in this work we opt for the second approach.


\section{Responsive Visualization Grammar}\label{sec:cicero}
\begin{figure*}[t]
    \centering
    \vspace*{5mm}
    \makebox[\linewidth]{
        \includegraphics[width=\textwidth]{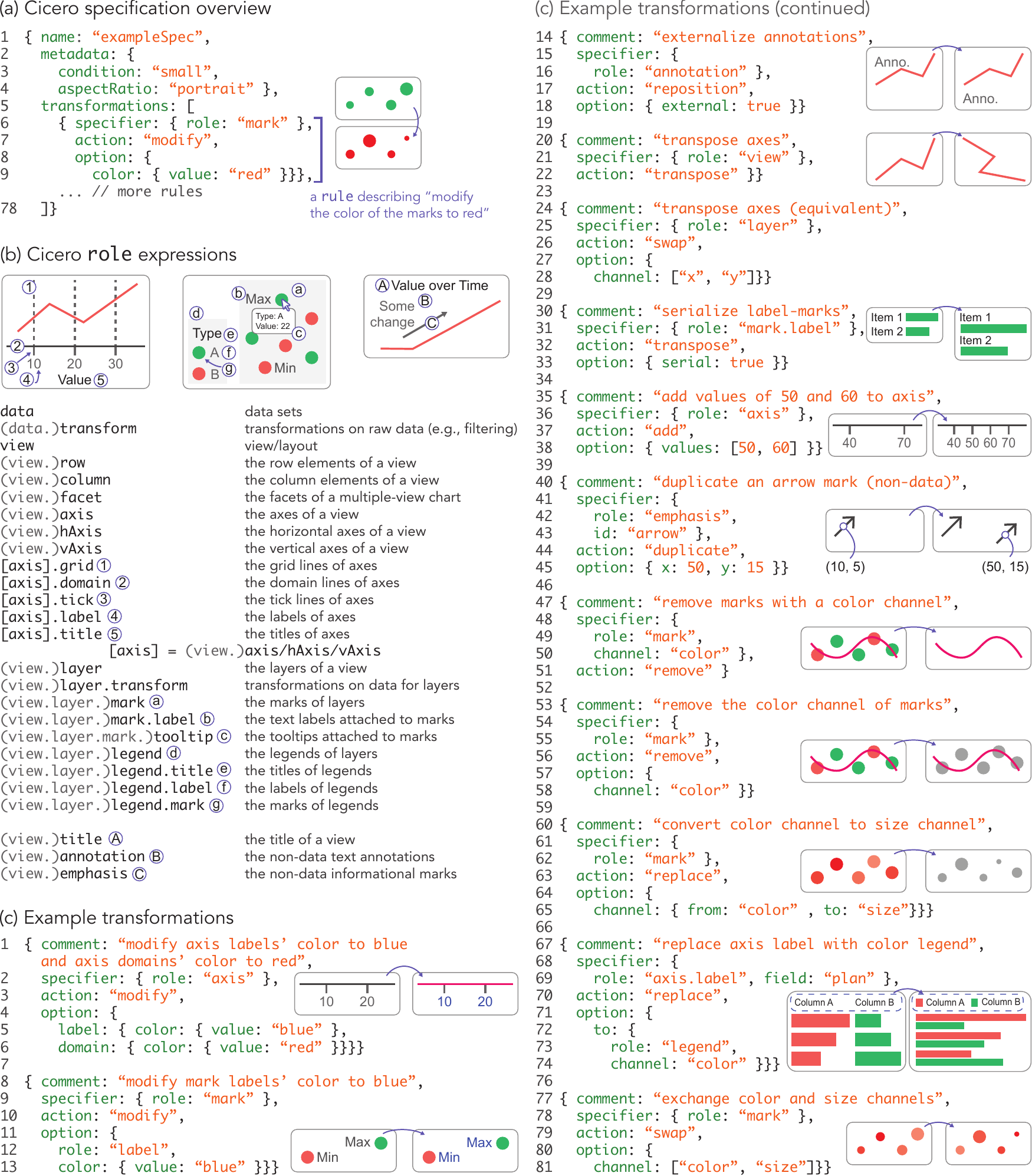}
    }
    \caption{Examples and roles in the Cicero grammar. (a) An overview of a Cicero specification with a rule describing ``\cicero{modify} the \cicero{color} of the \cicero{mark}s to \cicero{red}''. (b) \cicero{role} expressions used in Cicero. (c) Example transformations referred to in \autoref{sec:cicero}.}
    \Description{The contents of this figure is described in the caption and text. This figure contains three Sections, A, B, and C. For Section A, Cicero specification overview, read Section 4's introduction. For Section B, Cicero role expressions, read Section 4.1. For Section C, Example transformations, read Section 4.2.}
    \vspace*{-10mm}
    \label{fig:cicero_overview}
\end{figure*}

\begin{figure*}[t]
    \centering
    \makebox[\linewidth]{
        \includegraphics[width=\textwidth]{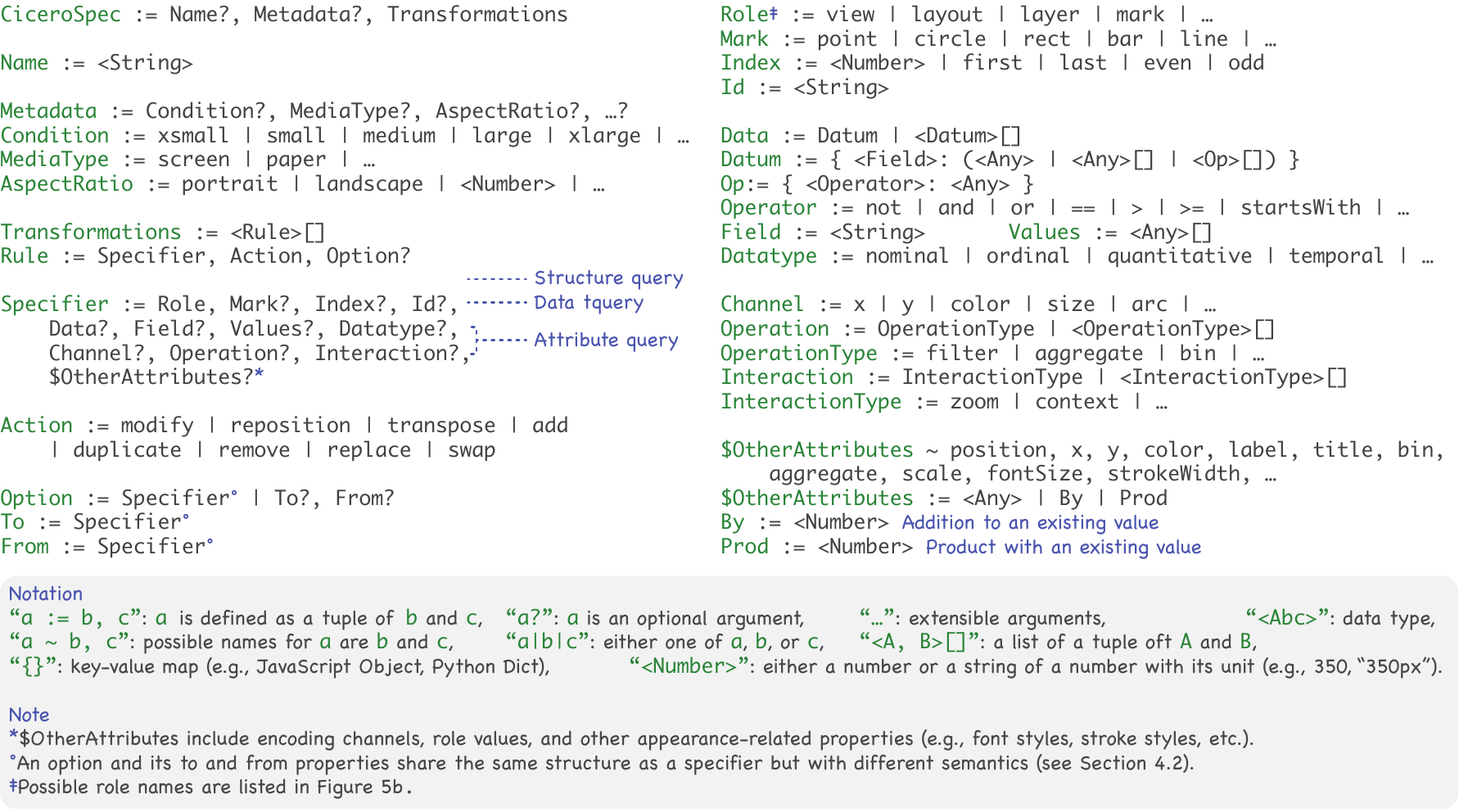}
    }
    \caption{The formal specification of Cicero. The Supplemental Material provides more detailed description.}
    \Description{A Cicero Spec object is a tuple of optional Name, optional Metadata, and required Transformations. Name is a string type element. Metadata is a tuple of optional Condition, optional Media Type, optional Aspect Ratio, and other option arguments that can be extended. Condition is either one of x-small, small, medium, large, x-large, or something else that can be extended. Media Type is either one of screen, paper or something else that can be extended. Aspect Ratio is either one of portrait, landscape, Number, or something else that can be extended. Footnote: Number means either a number or a string of a number and its unit. For example, 350 or 350 px. Transformations is an array of Rule. A Rule is a tuple of required Specifier, required Action, and optional Option. A Specifier is a tuple of required Role, optional Mark, Index, Id, Data, Field, Values, Datatype, Channel, Operation, Interaction, and dollar-sign-Other-Attributes. Role, Mark, Index, and Id comprise a structure query. Data, Field, Values, and Datatype comprise a data query. Channel, Operation, Interaction, and dollar-sign-Other-Attributes comprise an attribute query. Footnote: dollar-sign-Other-Attributes include encoding channels, role values, and other appearance-related properties, such as font styles, stroke styles, etc. Action is either one of modify, reposition, transpose, add, duplicate, remove, replace, and swap. Option is either one of a tuple of a required Specifier or a tuple of optional To and optional From. To is a tuple of a Specifier. From is a tuple of a Specifier. Footnote: An option and its To and From properties share the same structure as a specifier but with different semantics (see Section 4.2). Role is either one of view, layout, layer, mark, or something else that can be extended. Footnote: Possible role names are listed in Figure 5b. Mark is either one of point, circle, rect, bar, line, or something else that can be extended. Index is either a Number, first, last, even, an odd. Id is a string. Data is either a tuple of Datum or an array of Datum. Datum is a key-value map where a key is Field and its value is either Any type, an array of Any type, or an array of Op. Op is a key-value map where a key is Operator and its value is Any type. Operator is either one of not, and, or, equals, greater than, greater than or equal to, starts-with, or something else that can be extended. Field is a string. Values is an array of Any type. Datatype is either one of nominal, ordinal, quantitative, temporal or something else that can be extended. Channel is either one of x, y, color, size, arc, or something else that can be extended. Operation is either Operation-Type or an array of Operation-Type Operation-Type is either one of filter, aggregate, bin, or something else that can be extended. Interaction is either Interaction-Type or an array of Interaction-Type. Interaction-Type is either one of zoom, context, or something else that can be extended. Possible names for dollar-sign-Other-Attributes are position, x, y, color, label, title, bin, aggregate, scale, font-size, stroke-width, and something else that can be extended. dollar-sign-Other-Attributes is either one of Any type, By, or Prod. By is a tuple of a number that means addition to an existing value. Prod is a tuple of a number that means product with an existing value.}
    \label{fig:cicero_formal}
\end{figure*}

\noindent We introduce \textit{Cicero}, a declarative grammar designed to concisely express responsive transformations.
Paired with a declarative specification for a source visualization, Cicero provides a concise syntax for describing responsive changes independent of the structure of the original visualization specification.
A single Cicero specification defines how to \emph{transform} an initial visualization design to a new design, thereby encoding the responsive transformations required to convert a visualization into a responsive version for a particular format.
A Cicero specification consists of a metadata object (\cicero{metadata}, line 2--4 of \autoref{fig:cicero_overview}a) and a list of transformation rules (\cicero{transformations}, line 5--78 of \autoref{fig:cicero_overview}a). 
The \cicero{metadata} object contains meta-information about the context for the target view (\ie~the intended environment, including information like the media type and screen size).
The responsive strategies are encoded as separate rules in the list of \cicero{transformations}.
We use a `list' structure to enhance the reusability of the grammar by ensuring that each \cicero{rule} modularly describes a single responsive change to the source view (\cRef{dc:reusable}: reusable).
The formal specification of the Cicero grammar is shown in \autoref{fig:cicero_formal} and the Supplemental Material.

The core components of a \cicero{rule} object include the \cicero{specifier} (which elements to change), an \cicero{action} (how to change the elements), and the \cicero{option} (what properties to change). 
The \cicero{specifier} queries the source visualization to identify the set of existing visualization elements to be transformed, and supports flexibly referencing visualization elements with varying levels of scope (\cRef{dc:flexible}: flexible).
Then, the \cicero{action} and \cicero{option} provide high-level direction and detailed information about the change to be made to the selected elements, respectively, together encoding a wide range of transformations to elements selected by the \cicero{specifier} (\cRef{dc:expressive}: expressive).
For example, the rule object in line 6--9 of \autoref{fig:cicero_overview}a states that the compiler should `modify' (\cicero{action}) the `mark' (\cicero{specifier})'s `color' to be `red' (\cicero{option}).

In \autoref{sec:examples}, we provide a complete walk-through of the ``Bond Yields'' example; twelve additional examples are available in the Supplemental Material.
We chose properties and values for the \cicero{specifier}, \cicero{action}, and \cicero{option} in a principled fashion based on these example use cases (\autoref{sec:examples}) and prior work~\cite{hoffswell:responsive2020,kim:responsive2021}.
As a Cicero specification is independent of the structure of the source visualization, Cicero's properties and values can be extended in the future as needed.


\subsection{Specifier: Selecting elements to transform}\label{sec:specifier}
\noindent A \cicero{specifier} indicates which elements to transform on the target visualization. 
A \cicero{specifier} should only express existing element(s) from the target view, which the compiler then uses to identify the corresponding element(s) and extract relevant properties. 
Authors tend to apply responsive transformations to groups of element(s) sharing the same role, such as axis labels, mark tooltips, or legend marks, as characterized in prior work~\cite{hoffswell:responsive2020,kim:responsive2021}.
In addition, authors may want to include transformations specific to some data features (\eg~mark labels for specific data points, the axis corresponding to a particular data field) and/or the visual attributes of the visualization element(s) (\eg~red-colored bars). 
To express visualization elements using different characteristics, one can declare a specifier by \textit{structure}, \textit{data}, and \textit{attribute} queries. 

\bstart{Structure query} Many declarative visualization grammars like ggplot2~\cite{wickham:ggplot22010}, Vega~\cite{satyanarayan:vega2016}, and ZingChart~\cite{zingchart} define roles for visualization elements (\eg~marks, axes).
Structure queries identify elements based on this role, and provide additional flexibility for selecting and grouping elements in different ways, regardless of how the original visualization specification define them (\cRef{dc:flexible}: flexible). 
Keywords for structure queries include \cicero{role}, \cicero{mark}, \cicero{index}, and \cicero{id}. 

The \cicero{role} keyword specifies the role of a visualization element (see  \autoref{fig:cicero_overview}b).
The \cicero{role} can be cascaded to specify subordinate elements like \cicero{"mark.label"} for labels associated with the visualization marks or \cicero{"legend.mark"} for legend marks.
For brevity, cascaded role keywords can be shortened when its parent role is unambiguous (\eg~\cicero{"layer.mark"} as \cicero{"mark"}; \cicero{"view.row"} as \cicero{"row"}, possible short forms are indicated as gray-colored and parenthesized in \autoref{fig:cicero_overview}b).
The \cicero{mark} keyword specifies the type of mark, which is useful when there are multiple mark types in a visualization.
One can include the \cicero{index} keyword to indicate the specific element to select from a group of related elements (\eg~\specin{\{role:}\linebreak\specin{"title", index: 1\}} selects the second title element). 
To indicate the first and last element, one can use \specin{"first"} or \specin{"last"} for the index value.
Using \specin{"even"} and \specin{"odd"} can express every other (even and odd) element, respectively.
The \cicero{id} keyword selects informational marks (\cicero{emphasis}) 
by their defined names or identifiers (\eg~line 43 in \autoref{fig:cicero_overview}).

\bstart{Data query} A data query can reference a subset of data (\cicero{data}), a data field (\cicero{field}), the type of a variable (\cicero{datatype}), and values for elements (\cicero{values}) to support varying level of customization in selecting visualization elements (\cRef{dc:expressive}: expressive).
For example, the specifier \specin{\{role: "mark", data: \{price: 30\}\}} selects all marks that encode a price value of 30. 
Likewise, the specifier \specin{\{role: "axis", field: "price"\}} expresses axes for the \datafield{price} field; \specin{\{role: "legend", datatype: "nominal"\}} selects legends for nominal data variables.
The \cicero{values} keyword expresses a subset of values for a reference element that is tied to a certain data field like axis and legend .
For instance, the specifier \specin{\{role:}\linebreak\specin{"axis.label", values: [30, 50]\}} indicates the labels of an axis that encode value of 30 or 50.
Similar to the \cicero{index} keyword for a structural query, one can use \specin{"even"} and \specin{"odd"} to specify every other (even and odd) value element.
In order to support more complex data queries, we also provide a set of logical (NOT, AND, OR), arithmetic ($=, \neq, \leq, \geq, \le, \ge$), and string operations (\cicero{regex} pattern, \cicero{startsWith}, \cicero{includes}, \cicero{endsWith}) that can be composed to further select and filter elements based on properties of the data (\cRef{dc:flexible}: flexible). 

\bstart{Attribute query} 
An attribute query references visualization elements based on their properties or attributes. The primary attribute query keywords for identifying properties of visualization elements are: \cicero{channel}, \cicero{operation}, and \cicero{interaction}.
The \cicero{channel} keyword indicates whether the element has a certain encoding channel.
For instance, the specifiers \specin{\{role: "layer", channel:}\linebreak\specin{"color"\}} and \specin{\{role: "legend", channel: "color"\}} indicate layers and legends with a color encoding channel, respectively.
The \cicero{operation} keyword captures the type of data transformation operations applied to the elements (\eg~filter, aggregate), and the \cicero{interaction} keyword expresses the type of interaction features (\eg~zoom, interactive filter).
Cicero also supports the use of style and position attribute keywords such as color, font size, orient, relative positions \etc (see \cicero{\$OtherAttributes} in \autoref{fig:cicero_formal}). 
For marks, those style attributes can be used to indicate mark properties (\eg static color value or color encoding channel).
For example, \specin{\{role:}\linebreak\specin{"mark", color: "red"\}} indicates red-colored marks. 

\subsection{Action \& Option: Applying transformations}\label{sec:options}
The \cicero{action} indicates how to change the elements queried by a specifier. 
We designed Cicero to provide a concise set of action predicates that can encode a large range of transformations (\cRef{dc:expressive}: expressive).
The actions currently supported by Cicero are: \cicero{modify}, \cicero{reposition}, \cicero{transpose}, \cicero{add}, \cicero{duplicate}, \cicero{remove}, \cicero{replace}, and \cicero{swap}, chosen based on prior work~\cite{hoffswell:responsive2020,kim:responsive2021}.
Our aim was to support a minimal set of action predicates from the prior work~\cite{hoffswell:responsive2020,kim:responsive2021}.
For example, reposition actions in Kim~\ea~\cite{kim:responsive2021} can be efficiently expressed with using a single `reposition' action and various option properties (\eg~\cicero{externalize} $\rightarrow$ \cicero{reposition} + \specin{external: true} and \cicero{fix} $\rightarrow$ \cicero{reposition} + \specin{fix: true}). 
The `modify' action can also express these changes to positions, yet having a single `reposition' keyword is likely simpler for authors to remember.
This smaller set of action predicates does not sacrifice much expressiveness, as shown in our diverse set of examples in \autoref{fig:cicero_overview}, \autoref{sec:examples}, and the Supplemental Material.

The \cicero{option} object in a rule further details the change indicated by the \cicero{action}. 
While the core structure of an \cicero{option} object is the same as a specifier, the structure and keywords vary based on the type of action.
Keywords used in an \cicero{option} object refer to the properties or subordinate elements of the elements that were identified by the \cicero{specifier} (\eg~axis labels are subordinate elements of an axis), so a compiler should interpret an \cicero{option} object with regard to the \cicero{specifier}.

For example, one can use the \cicero{role} keyword to specify subordinate elements in an \cicero{option} object.
An \cicero{option} \specin{\{role: "label"\}} means legend labels if the specifier is \specin{\{role: "legend"\}} or mark labels if the specifier is \specin{\{role: "mark"\}}. 
When an \cicero{option} does not include the \cicero{role} keyword, then the properties in the \cicero{option} indicate those of the element identified by the \cicero{specifier}.
For example, in line 8--9 of \autoref{fig:cicero_overview}a, \cicero{"color"} refers to the color of the \cicero{"mark"} (the \cicero{specifier} in line 6),
while the \cicero{color} keyword in line 13 of \autoref{fig:cicero_overview}c expresses the color property of the marks' (\cicero{specifier}) labels (\cicero{option}).
Finally, when role values are used as a keyword in the \cicero{option}, they indicate the subordinate elements of the element specified by the \cicero{specifier}.
For instance, in \autoref{fig:cicero_overview}c, line 5--6 mean the color of the axes' (\cicero{specifier}) labels and domain lines (\cicero{option}), respectively.
The entire transformation rule (line 1--6) states that the compiler should specify all the axes in the chart, and modify the labels' color to be blue and the domains' to be red.

\bstartnc{A \cicero{modify} action} changes the properties of an element to specific values, with an associated \cicero{option} object for expressing attributes of the elements selected by the \cicero{specifier}.
For instance, one can modify the color of mark labels using the rule in line 8--13 of \autoref{fig:cicero_overview}.
To make relative changes, including adding or multiplying an attribute value by some value, one can use \cicero{by} and \cicero{prod} operators, respectively. 
For instance, a user can express modifying the size of the specified marks by subtracting 30 using the \cicero{by} operator: \specin{\{specifier: \{role: "mark"\}, action:}\linebreak\specin{"modify", option: \{size: \{by: -30\}\}\}}. 

\bstartnc{A \cicero{reposition} action} is a special type of the \cicero{modify} action designed to more intuitively support common transformations related to position properties like absolute positions (\cicero{x}, \cicero{y}), relative positions (\cicero{dx}, \cicero{dy}), externalization (\cicero{external}, \cicero{internal}), \etc 
For example, externalizing text annotations can be expressed as line 14--18 in \autoref{fig:cicero_overview}c.
If a user wants to change the style and position properties together, then a \cicero{modify} action is recommended. 

\bstartnc{A \cicero{transpose} action} expresses the relative position of a pair of elements, the relationship of which is defined a priori, like two positional axes (\textit{x} and \textit{y}), labels associated with an axis or marks.
A \cicero{transpose} action helps simplify expressions for relational properties.
For example, the rule in line 20--22 (\autoref{fig:cicero_overview}) transposes the entire channel. 
The equivalent is to \cicero{swap} the \textit{x} and \textit{y} position channels in \cicero{layer}s, as in line 24--28.
To serialize labels to their marks, one can use the rule in line 30--33 with a \cicero{serial} keyword in the \cicero{option}.
This behavior is the same as adjusting the label positions (relative \textit{x} and \textit{y} values) and mark offsets.

\bstartnc{An \cicero{add} action} adds new elements in a visualization. 
Since the \cicero{specifier} only expresses existing elements (\autoref{sec:specifier}), the newly added elements are provided in an \cicero{option} object. 
For example, to express ``\cicero{add} \cicero{values} of \cicero{50} and \cicero{60} to \cicero{axis}'', one can use the rule in line 35--38 in \autoref{fig:cicero_overview}c.
When the existing axis selected by the \cicero{specifier} (line 36) has ticks and labels for each axis value, 
then the rule should result in adding ticks and labels for those values specified in the \cicero{option} (line 38).

\bstartnc{A \cicero{duplicate}} action copies the element identified by the \cicero{specifier}.
If provided, an \cicero{option} indicates the properties for the duplicated element to change after duplication (\eg~repositioning the duplicated element in line 40--45 of \autoref{fig:cicero_overview}c). In this case, the \cicero{option} acts as a shortcut for a second \cicero{modify} transformation to update the newly added element.

\bstartnc{A \cicero{remove} action} removes elements identified by the \cicero{specifier} when no \cicero{option} is provided; when included, the \cicero{option} specifies the properties or subordinate elements that should be removed from the elements identified by the \cicero{specifier}. 
For instance, line 47--51 of \autoref{fig:cicero_overview}c removes all marks that include a \cicero{color} \cicero{channel} (no \cicero{option} is provided); to instead remove the color channel of these marks requires an \cicero{option} to be expressed (line 53--58).

\bstartnc{A \cicero{replace} action} expresses changes to the function of an entity while retaining its attributes.
Sometimes, a visualization author may wish to change the role of an element such as changing from axis labels to legends (\autoref{fig:flexibility}) or changing an encoding channel of the marks to use increased screen space efficiently.
There are two types of \cicero{replace} actions: replacing a property with another within an element and replacing the role of an element with another. 
For the first case, users can use the \cicero{from} and \cicero{to} keyword to indicate the original property and the replacing property. 
For instance, converting a channel from color to size can be expressed as the rule in line 60--65 (\autoref{fig:cicero_overview}c).
Second, authors often change the role of elements across the visualization structure, which requires an \cicero{option} to not be subordinate to the specifier. 
In that case, users can use a \cicero{to} keyword to indicate that this rule is changing the structural property. For instance, one can replace an axis for the field plan with a legend for the color channel (which is meaningful only when the color channel encodes the same field) by having a rule shown in line 67--74.

\bstartnc{A \cicero{swap} action} exchanges two entities (roles and encoding channels)
while retaining their properties, which shortens two replace actions and helps avoid potential conflicts. 
While a \cicero{swap} action has the same option structure with a \cicero{replace} action, it can also use an array to indicate properties to be swapped.
For instance, to exchange the color and size channels, one can have a \cicero{swap} action and an array-based \cicero{option} as shown in line 77--81 (\autoref{fig:cicero_overview}c). 

\subsection{Reusability of Cicero Expressions}
\begin{figure}[t]
    \centering
    \includegraphics[width=\columnwidth]{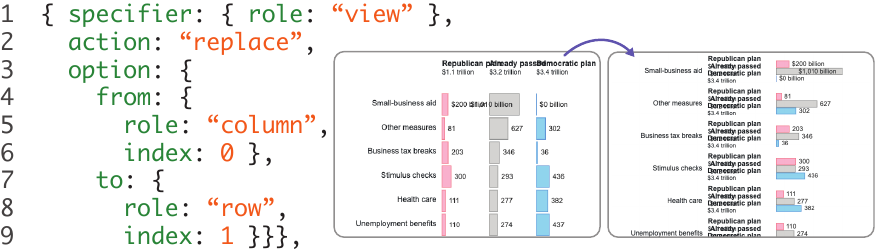}
    \caption{An example Cicero rule describing partial transpose. The bars are grouped by columns in the left view (before) and by rows in the right view (after). The entire set of transformations for this case (Aid Budget) can be found in the Supplemental Material.}
    \Description{This figure contains code snippets and two relevant demonstration images. The code snippets are described in the caption. The first demonstration image has a column-wise grouped bar chart, and the second image has a row-wise grouped bar chart.}
    \label{fig:reusability_example}
\end{figure}

Responsive transformation strategies differ in how well they generalize across visualizations. 
Sets of public-facing Web visualizations often appear together in a data-driven article and may share data sets, chart types, and style schemes, thereby facilitating transformation reuse.
For example, the data filtering rule in line 5--10 of \autoref{fig:walkthrough} can be reused for other charts sharing the same data set because it references the data fields (\cicero{year}, \cicero{forecasted\_year}) directly.
However, this rule cannot necessarily be reused on charts with different data sets. 
On the other hand, authors can reuse the partial axes transpose rule in \autoref{fig:reusability_example} for charts with a similar format regardless of the underlying data set as the transformation is declared independently.
The flexible specifier syntax of Cicero is designed to allow authors to express more reusable transformations.
For instance, the transformation for adding axis values in line 35--38 of \autoref{fig:cicero_overview}c can be reused on neighboring charts to provide better consistency.
Alternatively, one can express the same rule as \specin{\{specifier: \{role: "vAxis"\}, action: "add",}\linebreak\specin{option: \{index: "odd"\}\}} to make the rule more generalizable by not making direct reference to the underlying data scheme.
Expression reusability is a core attribute of Cicero that naturally supports sophisticated visualization authoring tools, such as recommender systems, which we discuss further in \autoref{sec:recommender}.


\section{Principles for Our Cicero Compiler}\label{sec:compiler}
\begin{figure}[t]
    \centering
    \includegraphics[width=\columnwidth]{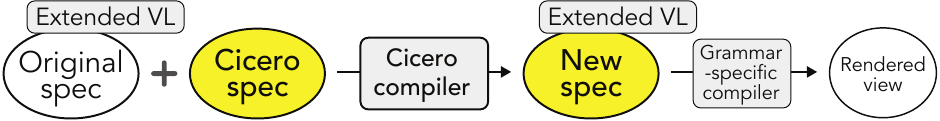}
    \caption{The pipeline for our Cicero compiler, developed for our extended version of Vega-Lite.}
    \Description{This diagram describes that an original spec in our extended Vega-Lite and a Cicero spec are passed to our Cicero compiler which produces a new spec in our extended Vega-Lite. This new spec is passed to the Grammar-specific compiler which returns a rendered view.}
    \label{fig:cicero_pipeline}
\end{figure}

\noindent To demonstrate the feasibility of Cicero and our proposed approach, we developed a compiler for our extended implementation of Vega-Lite. In the process, we identified ten principles we considered when implementing our Cicero compiler.
As outlined in \autoref{fig:cicero_pipeline},our prototype Cicero compiler takes as input a Cicero specification and a visualization design specification written in our extended Vega-Lite. 
Then, the Cicero compiler returns a transformed design specification in our extended Vega-Lite, which is eventually rendered by the compiler of our extended Vega-Lite.
For each \cicero{transformation} rule, our compiler first selects an element(s) indicated by the \cicero{specifier}. If the element(s) exists, then the compiler applies the changes specified by the \cicero{action} and \cicero{option}.
While developing the prototype compiler and deriving the principles below, we examined examples from prior work~\cite{hoffswell:responsive2020,kim:responsive2021} and considered how our compiler should deal with downstream effects to associated elements, the default behavior of a rendering grammar, and conflicting transformation rules.
Future work can leverage our principles as useful semantics of the Cicero grammar when implementing custom Cicero compilers for other declarative visualization grammars.
We describe our custom Cicero compiler API in the Supplemental Material (\url{https://osf.io/eg4xq}).


\bstartnc{Our extended version of Vega-Lite} provides a set of workarounds for public-facing visualization technique, such as text-wrapping and supplemental text (captions), that are currently not supported in Vega-Lite~\cite{satyanarayan:vega-lite2017}, but were needed for our examples (\eg~external annotations).
We use this extension to demonstrate the capabilities of Cicero for real-world use cases.
The key differences from the current Vega-Lite are that our extension (1) uses trellis plot-based layouts~\cite{becker:trellis1996} (rows and columns) instead of \textit{x} and \textit{y} encodings, (2) has many shortcuts to design techniques (\eg~wrapping text, map visualizations, interactive filters) for which Vega-Lite currently requires further specifications,
and (3) supports richer communicative functionalities such as defining supplemental text elements like multiple subtitles or captions, creating graphical emphases that are not bound to data, allowing different formats of labels in the same axis, and so forth.
The formal specification and description of our extended Vega-Lite are in the Supplemental Material.

\subsection{Associated elements}\label{sec:association}
Visualization elements can have associations between them, which should inform how our Cicero compiler selects and handles the elements. 
For example, axis labels are dependent on the range of visualized data encoded by the \textit{x} and \textit{y} positions; hence, axis labels are associated with the ranges of visualized data values (line 15--21 of \autoref{fig:walkthrough}).
When a subset of data is omitted under a responsive transformation, then text annotations attached to the corresponding marks should be omitted as well (line 5--10 of \autoref{fig:walkthrough}).

We describe two principles involving associated elements. First, our Cicero compiler  \principle{detects associated elements depending on how a user has defined the original design}{p:detection}.
In the previous example (\autoref{fig:walkthrough}), the two longer labels are declared as \texttt{text} elements of the line marks (\ie~tied to the marks in the same layer;\linebreak\specin{\{type: "on-mark",} \specin{field: "forecasted\_year",} \specin{items: }\linebreak\specin{[...], ...\}}). 
Thus, filtering out a subset of data subsequently removes the corresponding marks and their associated labels.
On the other hand, if the user has declared the text elements directly (without anchoring to certain data points), then the compiler should interpret them as independent elements that are not subordinate to any other element(s) or data.

\begin{figure}[t]
    \centering
    \includegraphics[width=\columnwidth]{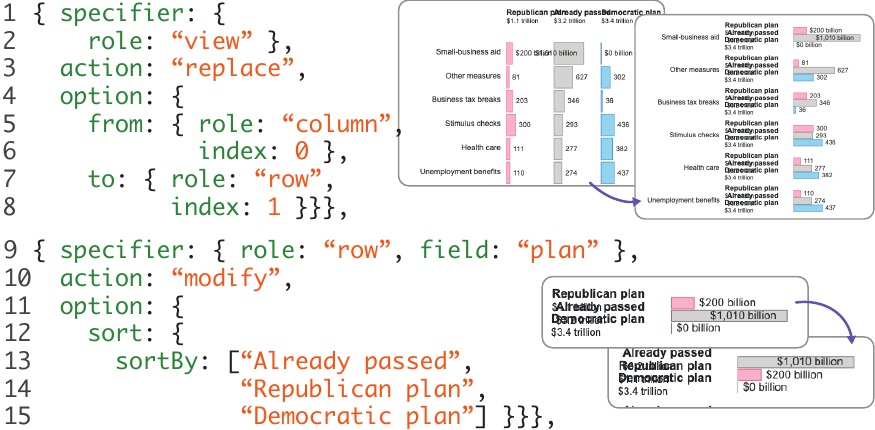}
    \caption{An example case (Aid Budget) for a downstream effect to the layout of elements (moving a column axis to a row axis; line 1--8) and applying a rule (reordering a nominal \textit{y} axis) to the previously transformed view (line 9--15).}
    \Description{This figure has code snippets and two sets of relevant demonstration images. The code snippets are described in the caption. The first image shows a column-wise grouped bar chart converted to a row-wise grouped bar chart with remaining axis labels. The second image shows the vertical order of bars in each row group from Republican plan, Already passed, and Democratic plan to Already passed, Republican plan, and Democratic plan.}
    \label{fig:compilerDownstream}
\end{figure}

Second, \textbf{a transformation affecting the layout of a series of elements}, such as adding, removing, or repositioning, \principle{has a downstream effect on the layout of their associated elements}{p:downstream}, but not the static style.
We do not allow downstream changes to style because the layout of one element and the static style of another are not meaningfully related whereas the relative layout between different elements does have a meaningful relationship.
In the previous example, filtering out data points should not impact any independent, non-data annotations but should remove any associated text element(s).
Similarly, converting a field from the column to the row of the chart (partial transpose) should move the axis labels (defined as \specin{\{type: "on-axis", field: "plan",}\linebreak\specin{items: [...], ...\}}; \ie~tied to the axis of the plan field) for the field accordingly (see line 1--8 in \autoref{fig:compilerDownstream}), but should not have side effects to their other properties---like the font weight or font size.


\subsection{Default behaviors}\label{sec:default_behavior}
Declarative grammars often have default behaviors to make it easier to create a visualization.
For example, Vega-Lite automatically generates legends and axis labels as a user declares color/size and position encoding channels.
In compiling a Cicero specification, we were able to relatively easily reason about default behaviors regarding removing, modifying, and externalizing actions (\eg~``modify only what is specified'' as a general software quality guideline or ``externalize annotations at the bottom of the chart unless specified otherwise'' based on our examples).
However, adding a new element and internalizing an element can complicate the compile process, particularly when a user has underspecified the behavior.
For example, when a user adds a new text annotation in the chart without specifying its position, then it is unclear how our Cicero compiler should behave.
To guide such complex situations, we used a set of high-level default behaviors for our Cicero compiler.

\begin{figure}[t]
    \centering
    \includegraphics[width=\columnwidth]{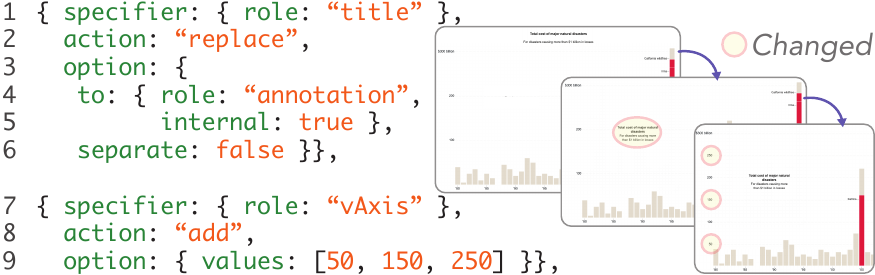}
    \caption{An example use case (Disaster Cost) for our Cicero compiler's default behavior for replacing the title as an internal annotation (line 1--6) and for introducing newly added elements (axis labels and grid lines; line 7--9).}
    \Description{This figure contains code snippets and three relevant demonstration images. The code snippets are described in the caption. The first demonstration image has a title on top of a bar chart. In the second image, the title is converted to an internal annotation located in an empty space of the chart. In the last image, three new vertical axis labels are added.}
    \label{fig:compilerDefault}
\end{figure}

First, \principle{when adding a new element to a series of elements, its appearance should mimic the existing elements in the series}{p:mimic}.
For example, line 7--9 in \autoref{fig:compilerDefault} adds new values for a vertical axis, resulting in newly added grid lines and labels.
Then, they should look similar to the existing grid lines and labels without further specifying their appearance.
Our Cicero compiler performs this addition by including those values in line 9 to the axis label and grid component in the specification (\ie~\specin{\{..., values:}\linebreak\specin{[100, 200, 300], ...\}} $\rightarrow$ \specin{\{..., values: [50, 100, 150, 200, 250, 300], ...\}}).


Second, our Cicero compiler \principle{considers the appearance of elements in a similar role for new elements that are not part of an existing series of elements}{p:assumption}.
For example, when adding labels to a \textit{y} axis that has no existing labels, although they are not in the same series, it is more sensible to set their appearance similar to the labels on the \textit{x} axis rather than the default presets of the rendering grammar.
The similarity of the role between two series can be determined by whether they can be specified as the same \cicero{role} keyword (\eg~\specin{\{role: "axis.label"\}} can specify both \specin{\{role:}\linebreak\specin{"hAxis.label"\}} and \specin{\{role: "vAxis.label"\}} if they both exist).
Then, our compiler reuses the appearance attributes of the similar series of elements.

\begin{figure}[t]
    \centering
    \includegraphics[width=\columnwidth]{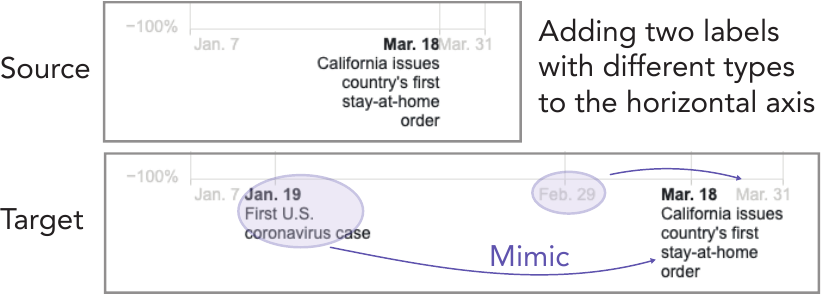}
    \caption{An example use case (Covid Spending 1) for our Cicero compiler's treatment of multiple series of existing elements. In this case, our Cicero compiler adds new axis labels by mimicking the most similar type of the existing axis labels according to the number of subelements (text lines).}
    \Description{This figure has a source and target design of a horizontal axis. In the source axis, there are three axis labels, where the middle has a date and description and the others have only dates. In the target axis, five axis labels are provided, where the second and third labels are marked as newly added. The second axis label has a date and description, marked as mimicking the fourth label, which is the middle one in the source design; and the third label has only a date, marked as mimicking the fifth label, which is the third one in the source design.}
    \label{fig:principle_mimic}
\end{figure}

Third, \principle{when there are multiple series of existing elements, our Cicero compiler selects the one with the most similar structure}{p:similarity}.
As shown in \autoref{fig:principle_mimic}, for instance, when adding a new label to an axis that has two groups of existing labels in different styles, our Cicero compiler reasons about which of the two groups is most similar to the new label.
We use the number of subelements (\eg~text segments) and the format of elements to find the most similar series of elements.
For our approach, the compiler first identifies the number of newly added text segments (two). 
The one starting with ``Jan. 19 ...'' has two segments with different styles, and the ``Feb. 29'' one has a single segment. 
Then, by comparing the numbers of segments, the compiler matches the two-segment one (``Jan. 19 ...'') with the new labels.


Lastly, we consider the case where the position and style of a newly added or repositioned element cannot be fully determined because there is no existing series with a similar role. 
In this case, the compiler should leverage the following default behavior if not specified otherwise: as an overarching principle, \principle{use the default options of the rendering grammar's compiler}{p:default} for newly added elements because users are expected to have some basic knowledge about how the rendering grammar behaves. 
For example, our extended Vega-Lite implementation does not automatically generate a legend for a new color scale, so our Cicero compiler for this extension similarly does not introduce a legend when adding a new color encoding.
On the other hand, Vega-Lite's default is to include a legend, so a Cicero compiler for Vega-Lite \emph{should} add a legend.
We had the following default behaviours for cases where the rendering grammar has no relevant default options based on our observations of common responsive design principles:

\begin{itemize}
    \item Place (new) externalized annotations below the chart (see a4 in \autoref{fig:recommender_example}).
    \item Place (new) internalized data annotations (or mark labels) at the center or the bottom of the associated data mark (see c4 in \autoref{fig:recommender_example}).
    \item Place (new) internalized non-data annotations at the center of the largest contiguous empty space in the chart (see line 1--6 in \autoref{fig:compilerDefault}).
\end{itemize}

\subsection{Conflict management}\label{sec:conflict}
Cicero's list-based specification explicitly indicates the order of declared transformation rules. 
However, there are some cases where the order of rules may impact how the Cicero compiler interprets a given specification.
Our compiler solves conflicts using the following methods, some of which are inspired by relevant CSS principles~\cite{MDN:cascade2020} that similarly deal with managing conflicts between ordered rule items. 
First, it may be confusing to select visualization element(s) in a \cicero{specifier} when other rules in the specification also transform the same element, which differs from general CSS use cases.
For example, suppose there is a rule to transpose the \textit{x} and \textit{y} positions.
This rule also results in swapping the horizontal and vertical axes as they are associated with the \textit{x} and \textit{y} position encoding channels.
If a user wants to make some design changes in an axis that is the horizontal axis after transposing but is the vertical axis before transposing, defining a specifier for this rule might be confusing.
A simple approach defaults to always specifying what is in the original view specification or what will appear in the transformed view.
However, the former may not be useful for cases like making further changes to a newly added element, and the latter might make it difficult to compose a specification by requiring users to imagine the outcome status. 
As an overarching principle, our compiler \principle{applies the current rule to a view that has been transformed by the previous rules}{p:order} (\eg~line 1--8 and line 9--15 in \autoref{fig:compilerDownstream}).
This approach also implies that the compiler \principle{applies the last declared rule}{p:last} when there are two rules making changes to the same element for the same property, which is also a common practice with CSS specifications. 
Our compiler handles this principle by updating the target view specification for each transformation rule.

\begin{figure}[t]
    \centering
    \includegraphics[width=\columnwidth]{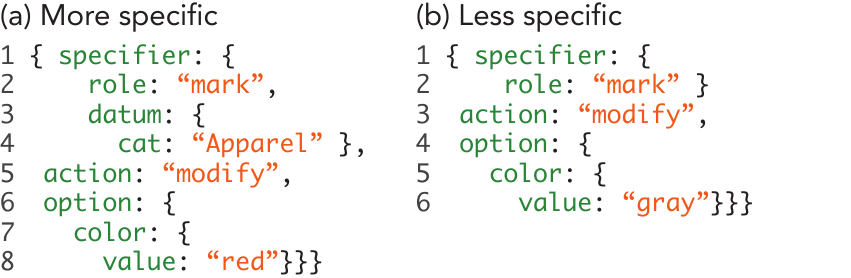}
    \caption{Rules to change the color of marks (a) by specifying the mark for the ``Apparel'' category and (b) by generally changing the color of all marks (independent of the data).}
    \Description{This figure has two code snippets. Snippet A on the left column has a specifier consisting of a role property of mark and a datum property of Apparel category, while Snippet B on the right column has a specifier only with a role property of mark.}
    \label{fig:specificity}
\end{figure}

Next, our compiler \principle{assigns higher priority to a more specific rule than a more generic rule for the same element}{p:specific} (note: not the same specifier)\footnote{See the `Justice Kennedy' case (desktop to mobile) in our Supplemental Material.}.
Here, the more attributes a \cicero{specifier} has, the more specific the rule is, inspired by CSS principles~\cite{MDN:specificity2020}.
For example, suppose a user wants to change the color of a mark for the ``Apparel''  category (rule (a) in \autoref{fig:specificity}) as well as changing the color of all bars (rule (b)).
Here, the mark for ``Apparel'' is affected by both rules.
Therefore, we recommend that generic color changes to other bars should not be applied to the mark for the ``Apparel'' category (\ie~rule (a) has higher priority than (b)).
If a user does not want to apply a specific change (\eg~the custom color for the ``Apparel'' mark), then the user should omit the rule from the Cicero specification.
Lastly, to enhance the degree of freedom in indicating the priority of rules, Cicero provides an \cicero{important} property for the same specifier, inspired by the \texttt{!important} keyword in CSS~\cite{MDN:specificity2020}.
\principle{Rules with the \cicero{important} property set to \cicero{true} have higher priorities than others}{p:important}
(\ie~compiled at the end).
For example, a rule that changes the color of every axis label with \specin{\{important: true\}} overrides another following rule that recolors a specific axis label\footnote{See the `Disaster Cost' case (desktop to mobile) in our Supplemental Material.}.
We refer the reader to the Supplemental Material for the full details on how our Cicero compiler for the extended version Vega-Lite exhibits these principles. 


\section{Reproducing Real-World Examples}\label{sec:examples}


\begin{figure}[t]
    \centering
    \includegraphics[width=\columnwidth]{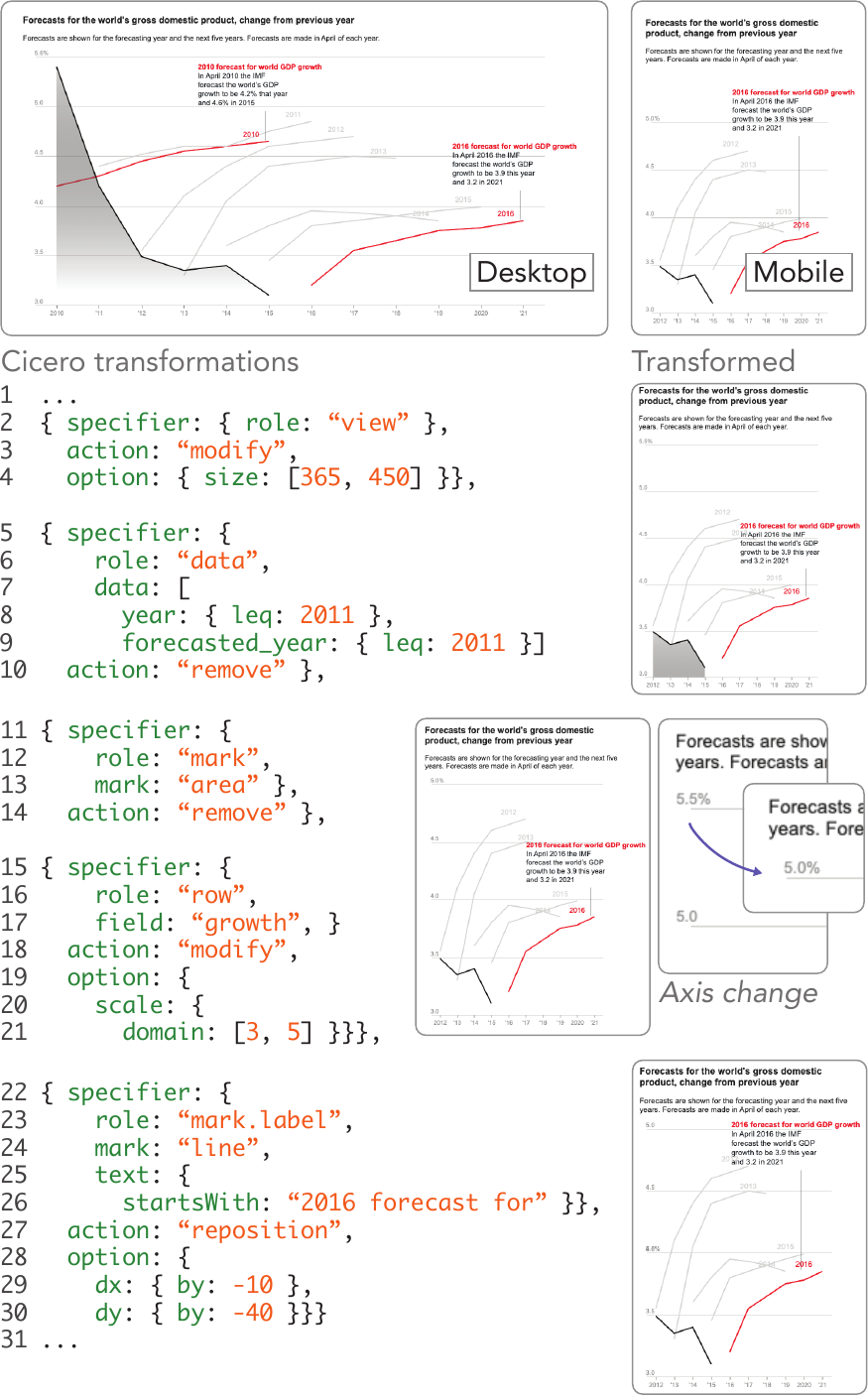}
    \caption{A walk-through example case of Bond Yields from a desktop version (top left) to a mobile version (top right). Starting with the desktop version, we first resize the chart to fit to a mobile screen (line 2--4), remove a subset of data for earlier years (line 5--10), remove the area mark (line 11--14), update grid lines by rescaling the domain of the \textit{y} position channel (line 15--21), and reposition the annotation (22--30).
    }
    \Description{This figure has two parts. The top part contains the desktop and mobile versions of a line graph. The bottom part consists of two columns. The left column has code snippets for the changes from the desktop to the mobile view. The right column has three relevant demonstration images for the corresponding code snippets. The contents of this figure is detailed in the text.}
    \label{fig:walkthrough}
\end{figure}

\noindent To demonstrate the expressiveness, flexibility, and reusability of Cicero and illustrate the above principles of our Cicero compiler, we present an in-depth walk-through of a mobile-to-desktop example (Bond Yields) using our extended version of Vega-Lite as the rendering grammar. 
We have twelve additional real-world inspired walk-through specifications that show the responsive changes step-by-step and two other detailed textual walk-throughs in the Supplemental Material that exhibit a variety of other transformations to visualization elements (\ie~data, marks, axes, title, labels, annotations, informational marks/emphasis, interaction, \etc) for both desktop-to-mobile and mobile-to-desktop transformations.
The total of 13 example use cases includes three from Hoffswell~\ea~\cite{hoffswell:responsive2020}, four cases from Kim~\ea~\cite{kim:responsive2021}, three recent responsive visualization cases (in our extended Vega-Lite), and two additional cases from the Vega-Lite example gallery that were not originally responsive but demonstrate the generalizability of our Cicero specifications to refine complex source views (in Vega-Lite).
All 13 cases are listed in \autoref{fig:teaser} and provided in the Supplemental Material (\url{https://osf.io/eg4xq}).

\subsection{A Walk-through Example: Bond Yields}\label{sec:bondyields}
\noindent The Bond Yields example\footnote{\url{https://www.wsj.com/graphics/how-bond-yields-got-this-low/}} visualizes changes to both the actual and forecasted GDP growth rates over time.
In the desktop version (\autoref{fig:walkthrough}), the \textit{x} position encodes the year from 2010 to 2021, and the \textit{y} position indicates the GDP growth rate from 3.0 to 5.5.
The area mark and black line mark represent the actual GDP growth rate from 2010 to 2015.
The red and gray lines represent the five-year forecast of GDP growth rate for each year from 2010 to 2016; for example, the leftmost red line shows the estimated GDP growth rates for 2011 to 2015, as forecast in 2010. 
Transformations to produce the mobile version include (1) reducing the chart size, (2) removing the data points and labels for the forecast year of 2010 and 2011, (3) omitting the area mark, (4) truncating the \textit{y} axis, and (5) repositioning an annotation. The Cicero spec is shown in \autoref{fig:walkthrough}a.

First, to resize the chart for a mobile phone, one can apply a \cicero{modify} action to the entire \cicero{view} (line 2--4).
The \cicero{option} object indicates the \cicero{size} of 365 (width) $\times$ 450 (height) to ensure that the chart fits a mobile phone without requiring horizontal scrolling.
Alternatively, one can use \specin{\{width: 365, height: 450\}} in the \cicero{option}.
Then, line 5--10 filters out (\cicero{remove}) the specified data points to simplify the view by reducing the information density.
The \cicero{data} keyword in the \cicero{specifier} means $\langle$year $
\leq$ 2011 (for the actual GDP growth rate) OR forecast year $\leq$ 2011 (for the forecast)$\rangle$.
Filtering out the data points removes (1) the two simple line marks for the forecast year of 2010 and 2011, (2) the data annotation for forecast year 2010, and (3) the corresponding parts of the area and black line mark for the actual GDP growth rates because each of these elements is associated with the filtered data (\pRef{p:downstream}).
This association is determined by the original visualization structure; if the annotations were declared as non-data elements, then the annotation for the 2010 forecast would remain (\pRef{p:detection}).

The \cicero{remove} transformation in line 11--14 omits the area mark specified by the \cicero{mark} keyword.
After filtering the earlier data, there is wasted space along the y-axis that unnecessarily compresses the data. 
To address this issue, the rule in line 15--21 changes the scale domain of the row field (\cicero{growth}) to \cicero{[3,5]}, resulting in the removal of the axis label and grid line for 5.5; the remaining elements automatically adjust to fill the newly vacated space (\pRef{p:default}).

Lastly, the \cicero{reposition} rule in line 22--30 moves the mark label.
Because there are many text elements associated with data marks (\eg~year names for each line), a specific \cicero{text} query is needed to select the label to move.
For this rule, one can use the \cicero{startsWith} operator (line 25--26) to select elements with text starting with the specified string.
Then, the \cicero{option} object changes the relative horizontal and vertical position (\cicero{dx} and \cicero{dy}, respectively) using the \cicero{by} operator which adds the specified value to the original value (\ie~moving the element by 10px left and by 40px upward).\\




\section{Potential Applications for Cicero}\label{sec:recommender}
Declarative grammars are particularly valuable for their utility in applications like visualization recommenders and authoring tools.
In particular, they can function as a common representation method for different intelligent tools with similar purposes~\cite{wu:ai4vis2021}.
Visualization systems often use their own ``internal representation'' methods for their specific purposes~\cite{wu:ai4vis2021}. 
Suppose we have two recommender models for different parts of a visualization (\eg~one for chart types and the other for annotations and emphases) that use heterogeneous representation methods. 
If they are translated to Cicero, then their recommender outcomes could be effectively combined to a user-side application.
In this section, we describe how we used Cicero to represent a design space of responsive transformations in a prototype design recommender for responsive visualization as a proof of concept. 
We further discuss how Cicero might support mixed-initiative authoring tools. 

\subsection{Responsive Visualization Recommender}
As a case study for potential applications for Cicero, we developed a recommender prototype for responsive visualization transformations using Answer Set Programming (ASP), which represents knowledge in terms of facts, rules, and constraints~\cite{brewka:asp2011}.
Our recommender takes a source visualization specification expressed in our extended version of Vega-Lite along with configuration preferences (\eg~intended screen size, strategies that a user wants to avoid, and a subset of data that can be omitted) which could hypothetically be provided by a user.
Our recommender is intended to provide a diverse set of recommendations rather than showing several ``optimal'' visualization with slight differences.
We encoded a set of common responsive visualization strategies motivated by prior work~\cite{hoffswell:responsive2020,kim:responsive2021} in ASP. 
Given the inputs and encoded strategies, Clingo~\cite{gebser:clingo2014}, an ASP solver, generates a search space of responsive transformation strategy sets (corresponding to responsive visualization designs).
To rank these strategy sets, we encoded heuristic-based costs that apply to individual strategies, and normalize and aggregate these costs to rank strategy sets representing design alternatives. 
We implemented three types of costs that apply to individual strategies: ``popularity'' costs based on the frequency of the strategy in prior analyses of professionally-designed responsive visualizations~\cite{hoffswell:responsive2020,kim:responsive2021}; 
density costs, where strategies that reduce information density are assigned lower cost than those that do not in a desktop-first pipeline, and vice versa in a mobile-first pipeline; 
and message preservation costs, where strategies (\eg~axis transpose, disproportional rescaling) are assigned costs based on the extent to which prior work proposes that they affect the implied ``message'' of a visualization~\cite{kim:insight2021,kim:responsive2021}.

In this pipeline, each recommended strategy set in the ASP format (\eg~\asp{do(transpose\_axes).}) are translated to a Cicero spec (\eg~\specin{\{specifier: \{role: "view"\}, action: "transpose"\}}). \\
While inference engines or models (\eg~ASP, ML, \etc) often employ their own abstract expressions for computational purposes, systems need to translate such abstract expressions (\eg~to JavaScript, Python, \etc) before utilizing them.
For instance, ASP can efficiently perform logic problems, but the ASP expressions cannot be directly used to execute actual tasks without translation.
In the context of responsive transformation, directly using ASP codes to transform a visualization design specification (\ie~running JavaScript codes for each ASP code) is likely to complicate the translation, lacking modularization.
For example, whenever a recommender adds a new transformation strategy, the system has to look at every detail of different use cases, and doing so may not be consistent with the existing transformation strategies.
This inconsistency in turn makes it more difficult to debug and extend the recommender.
Instead, if we can translate those abstract transformations to systematic expressions like the Cicero grammar, then implementing recommenders for responsive visualization only needs to focus on generating a search space by modularizing the translation process.
This process is similar to how Draco translates ASP expressions to Vega-Lite~\cite{satyanarayan:vega-lite2017} and then renders a visualization~\cite{moritz:draco2019}.

\begin{figure}[t]
    \centering
    \includegraphics[width=\columnwidth]{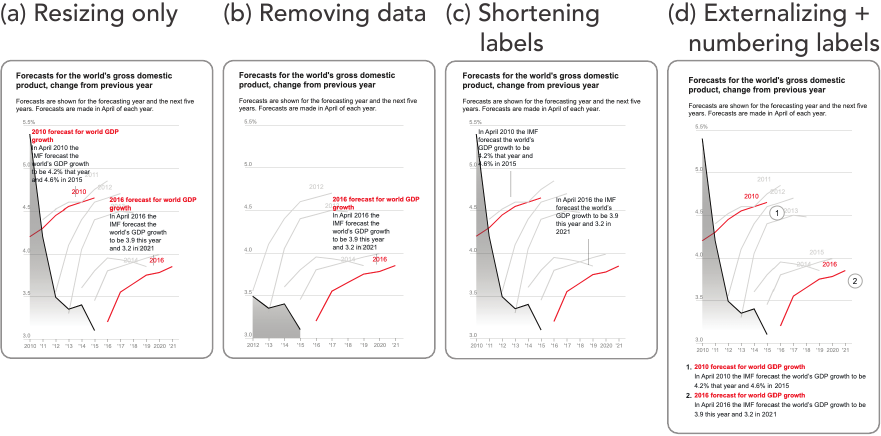}
    \caption{Selected examples among top seven recommendations for Bond Yields case from desktop to mobile. The original design is shown in \autoref{fig:walkthrough}.}
    \Description{This figure consists of for images of recommended versions for a line graph. The first image, a1, is simply resized to a mobile screen. The second image, a2, is resized and a part of data is removed. The third image, a3, is resized and parts of text annotations are removed. The last image, a4, is resized, and text annotations are externalized to the bottom of the chart.}
    \label{fig:recommender_example}
\end{figure}

Below, we illustrate example recommendations (\autoref{fig:recommender_example}a) using our walk-through example (\autoref{sec:examples}).
We provide further details on our prototype recommender implementation, and describe example recommendation cases below and in Supplemental Material (\url{https://osf.io/eg4xq}). 
We emphasize, however, that our goal in developing the prototype recommender is to demonstrate the feasibility of using Cicero in such an approach, rather than to argue for the specific implementation of the cost model we used. 
In other words, our recommender should be interpreted as a proof of concept of our approach, rather than as an ideal recommender.


\subsubsection{Example: Bond Yields}
To generate candidate mobile views for the Bond Yields case, we include in the configuration the target size of a mobile view and a subset of data that can be omitted (referring to the original design).
The first recommendation (\autoref{fig:recommender_example}a) is simply resized to the target size. 
For this change, our ASP recommender returns \asp{do(set\_width,365).} and \asp{do(set\_height,450).}, and these abstract descriptions are translated to corresponding Cicero rules: \specin{\{specifier: \{role: "view"\}, action: "modify",}\linebreak\specin{option: \{width: 365, height: 450\}\}}.
In the second recommendation (b), the suggested omission is applied, similar to the original mobile view except for the remaining area mark and axis value for 5.5\%.
Our ASP engine expresses the transformation in an abstract way (\asp{do(add\_filter,f0).}, where \asp{f0} is a pointer to the user-suggested data filter statement), and then it is converted to a proper Cicero rule, \specin{\{specifier: \{role: "data", data:}\linebreak\specin{[...]\}, action: "remove"\}}.
The data annotations for the forecast years of 2010 and 2016 are shortened by removing the first line (the red text) in the third recommendation (c).
For this change, our recommender converts an ASP rule, \asp{do(remove\_text\_line,t2,0).} where \asp{t2} is a pointer to the annotations (or mark labels), to a Cicero rule:
\specin{\{specifier: \{role: "mark.label", field:} \linebreak\specin{"forecasted_year",} \specin{index: 2\},} \specin{action: "remove",}\linebreak\specin{option: \{items: \{index: 0\}\}\}}.
The fourth recommendation (d) externalizes the same data annotations below the chart with numbering for reference to the data marks.
For this transformation, ASP rules, \asp{do(externalize,t2).} and \asp{do(numbering,t2).}, are translated to a Cicero rule: \specin{\{specifier: \{role: "mark.label",}\linebreak\specin{field: "forecasted_year", index: 2\}, action: "modify",}\specin{option: \{external: true, number: true\}\}}.
If the ASP rules were not compiled into our modularized Cicero grammar, the required changes to the original visualization specification would need to directly dissect many different parts of the specification, such as data, annotations, and axes.
By modularizing this computation, Cicero can provide a more systematic representation of those changes, which helps extend and debug our recommender.

\subsubsection{Generalizability for Recommenders}
\begin{figure}[t]
    \centering
    \includegraphics[width=\columnwidth]{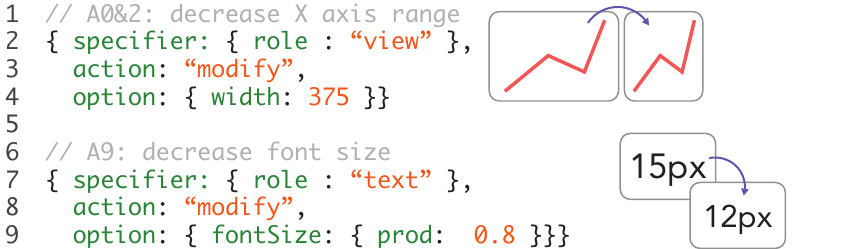}
    \caption{Expressing transformation strategies of MobileVisFixer~\cite{wu:mobilevisfixer2020} in Cicero. Line 2--4: decreasing the range of the \textit{x} axis by reducing the width of the chart. Line 7--9: decreasing the font size using \cicero{prod} keyword. 
    }
    \Description{This figure has two code snippets and the relevant demonstration images. The first code snippet is a Cicero rule modifying the view's width to 375, and the image shows a line chart converted to the resized version. The second code snippet is a Cicero rule modifying the font sizes of all the text elements by eighty percent, and the image shows a text element in the font size of fifteen pixels converted to a text element in the font size of twelve pixels.}
    \label{fig:generalizability}
\end{figure}

Cicero can enhance modularization of responsive visualization tools by connecting tool-specific expressions and visualization grammars.
For example, our recommender prototype uses ASP~\cite{brewka:asp2011} to encode expressions with the Clingo solver~\cite{gebser:clingo2014}), and the Cicero compiler connects recommendations expressed in ASP to visualizations in our extended Vega-Lite.
Future work might start to leverage Cicero with machine learning-based recommenders.
For instance, Cicero can express reusable transformation rules in MobileVisFixer~\cite{wu:mobilevisfixer2020} that translates non-responsively designed visualizations to mobile views.
As shown in line 2--4 of \autoref{fig:generalizability}, Cicero expresses `reducing the range of \textit{x} axis' by expressing the change to the chart width (\eg~375 pixel for mobile screens). 
Using the \cicero{prod} keyword in line 9, one can express reducing the font size of all the text elements relatively.
In the Supplemental Material, we provide a list of reusable Cicero expressions for MobileVisFixer~\cite{wu:mobilevisfixer2020} rules of which the meanings are clearly defined.


\subsection{Mixed-initiative Authoring Tools}
Users of visualization authoring tools may prefer different levels of customization and automation~\cite{mendez:tradeoff2017}.
Tools like Microsoft Power BI~\cite{gal:powerbi2017}, which automates design recommendations by converting a source visualizations using a set of default strategies, allow quick visualization creation, but can limit design expressiveness.
In contrast, while the prototype proposed by Hoffswell~\ea~\cite{hoffswell:responsive2020} and DataWrapper~\cite{datawrapper} do not have automated recommendation features, they enable more customization in making responsive designs.

Mixed-initiative authoring tools can provide a balance of automation and customization capabilities, by allowing authors the ability to make manual responsive transformations or accept recom\-men\-der-suggested transformations.
Mixed-initiative authoring has been applied in exploratory data analysis (\eg~Voyager~\cite{wongsuphasawat:voyager2016} and Dziban~\cite{lin:dziban2020}) and dashboard design (\eg~LADV~\cite{ma:ladv2021}) settings.
While our prototype recommender takes as input a representation of users' preferences, a next-generation authoring tool might aim to reason about responsive transformations that the user makes so as to recommend further or alternative transformations.
For example, imagine creating the Bond Yields case (\autoref{sec:bondyields}) without data filtering.
After resizing, the target visualization might look dense (\autoref{fig:recommender_example}a1) although it maintains more takeaways compared to the actual design.
Then, a user might decide to externalize the annotations instead of removing data.
Following this manual change, a mixed-initiative authoring tool might suggest numbering the externalized annotations to support finding data references.

A mixed-initiative approach stands to reduce computational complexity by looking at the current state of edits rather than reasoning over a larger space of transformation combinations. 
Within a mixed-initiative authoring pipeline for responsive visualization, Cicero can be used to represent both system-recommended transformation strategies and user-driven manual edits, which can make such systems easier and more efficient to handle different sources of transformations (system and user).
In addition, when an author updates the source visualization, Cicero can be used to reapply previous rules that are generalized to the updated chart (\ie~rules with the specifiers that can make queries from the updated chart).



\section{Limitations}\label{sec:limitations}
While Cicero and the Cicero compiler for our extended version of Vega-Lite can reproduce real-world use cases that represent a diverse set of transformations, future work should apply Cicero and future Cicero compilers to a bigger set of use cases to improve them and further extend the expressiveness of the grammar.
For example, future work might focus on expressing complex user interactions (\eg~pan+zoom for a 3D visualization) with \cicero{specifier}s, inspired by declarative grammars for interactive visualizations (\eg~\texttt{trigger}, \texttt{signal}, and \texttt{event streams} in Vega~\cite{satyanarayan:vega2016,vega:doc}), to better facilitate the application of such technologies to Web contexts where they have largely been underutilized~\cite{hoffswell:responsive2020,kim:responsive2021}.
Another interesting future direction could be expressions for bounded dynamic behavior---the sizes or arrangement of elements dynamically change up to a certain limit, such as \texttt{max-width} and \texttt{flex-wrap} in CSS---in \cicero{option}s.
As it is a Web browser that implements CSS specifications, additional expressions for bounded dynamic behavior will be useful only if a rendering grammar supports such behavior.
Furthermore, new design and evaluation studies for intelligent responsive authoring tools with Cicero might be useful to extend both Cicero and prior approaches in responsive visualization tooling~\cite{hoffswell:responsive2020,kim:insight2021,kim:responsive2021,wu:autolayout2021,wu:mobilevisfixer2020,leclaire:r3sjs2015,zingchart,rost:datawrapperAnno2020}.

Next, to demonstrate the full potential of Cicero in Web-based communicative visualizations, we chose to implement an extended version of Vega-Lite that can more easily express common techniques for narrative visualizations, such as externalizing annotations and applying word wrap to text labels.
These capabilities are not straightforward to implement in Vega-Lite~\cite{hoffswell:responsive2020}, so the resulting capabilities of a Cicero compiler for Vega-Lite may likewise be limited in what can be expressed in rendered visualizations.
As such grammars continue to develop, the corresponding compiler can be refined to support additional responsive functionalities.
Furthermore, future work might need to apply these techniques to a larger class of declarative systems, such as extensions based on ggplot2~\cite{wickham:ggplot22010} or Vega~\cite{satyanarayan:vega2016}, to efficiently implement the corresponding Cicero compilers with a better understanding of their capabilities.

Finally, a Cicero specification defines a set of transformations to create a single responsive version and itself is not intended for direct rendering. 
As multiple responsive versions are necessary for different device types, an authoring system could bundle multiple Cicero specifications as a family using the \cicero{metadata} object in the specifications to decide when to apply each of them.

\section{Conclusion}
We contribute Cicero, a declarative grammar for specifying responsive transformations from a source to a target visualization. By enabling flexible, expressive, and reusable specifications of visualization transformations, 
Cicero paves the way for intelligent responsive visualization authoring tools,
by providing a concise set of action predicates that enable encoding diverse transformations, flexible specifier syntax for handling the behavior of transformations, and reusability of transformation rules.
To demonstrate the utility of Cicero in the context of intelligent visualization tools, we leverage Cicero for a prototype design recommender for responsive transformations.
Future work can employ Cicero for a range of responsive visualization authoring tools designed for specific declarative grammars with custom compilers for those grammars.



\balance
\bibliographystyle{ACM-Reference-Format}
\bibliography{bibliography}

\end{document}